\documentclass{aa} 
\usepackage{graphicx}
\usepackage{float}  
\usepackage{pdflscape}
\usepackage{txfonts}
\usepackage{xcolor}
\usepackage{placeins}
\usepackage{booktabs}
\usepackage{multirow}
\usepackage[table]{xcolor}
\usepackage{subcaption}
\usepackage{makecell}
\usepackage{graphicx}
\usepackage{subcaption}  % Make sure to include this in your preamble
\renewcommand{\arraystretch}{1.2}

\begin{document} 

\title{RGC: a radio AGN classifier based on deep learning}
\subtitle{I. A semi-supervised multiclass model for VLA images}
\author{
M. S. Hossain\inst{1} \and
M. S. H. Shahal\inst{2} \and
K. M. B. Asad\inst{2,3} \and
P. Saikia\inst{5} \and
A. Khan\inst{1} \and
F. Akter\inst{6} \and
A. A. Ali\inst{1,4} \and
M. A. Amin\inst{1,4} \and
D. P. Guha\inst{2} \and
M. O. B. Jihad\inst{2} \and
A. Momen\inst{1,2,3} \and
S. Sen\inst{2} \and
A. K. M. M. Rahman\inst{1,4}
}
\institute{
Center for Computational and Data Sciences, Independent University, Bangladesh, Dhaka 1229, Bangladesh\\
\email{sazzat@iub.edu.bd} \and
Center for Astronomy, Space Science and Astrophysics, Independent University, Bangladesh, Dhaka 1229, Bangladesh\\
\email{kasad@iub.edu.bd} \and 
Department of Physical Sciences, Independent University, Bangladesh, Dhaka 1229, Bangladesh \and 
Department of Computer Science and Engineering, Independent University, Bangladesh, Dhaka 1229, Bangladesh \and 
Department of Astronomy and Physics, Yale University, New Haven, CT 06511, USA \and
Department of Agricultural and Biosystems Engineering, North Dakota State University, Fargo, ND 58108, USA
}

\date{Received ; accepted }

% \abstract{}{}{}{}{} 
% 5 {} token are mandatory
 
  %\abstract
  % context heading (optional)
  % {} leave it empty if necessary  
   %{con}
  % aims heading (mandatory)
   %{}
  % methods heading (mandatory)
   %{}
  % results heading (mandatory)
   %{}
  % conclusions heading (optional), leave it empty if necessary 
   %{}

\abstract
  % context
   {Bent radio active galactic nuclei (RAGNs)---wide-angle tails (WATs) and narrow-angle tails (NATs)---trace dense environments in galaxy groups and clusters, yet no multiclass classifier simultaneously separates them from straight Fanaroff--Riley types (sFRI, sFRII) using visually inspected labels and unlabelled data.}
  % aims
   {We release FIRST-2060, a four-class labelled dataset of 2060 RAGNs (sFRI, sFRII, WAT, NAT) constructed from three publicly available catalogues through multi-tier visual inspection, together with the semi-supervised RGC 1.0 model that leverages 20\,000 unlabelled radio sources. We benchmark RGC 1.0 against five supervised baselines spanning CNNs and vision transformers.}
  % methods
   {FIRST-2060 is provided in two preprocessing variants: $\mathbf{R}_{L1}$, which retains spurious sources, and $\mathbf{R}_{L2}$, from which they are removed. The RGC model integrates the self-supervised framework BYOL (Bootstrap Your Own Latent) with an $E(2)$-equivariant steerable CNN (E2CNN) encoder, pre-trained on the unlabelled data and fine-tuned on the labelled sets. All six models are evaluated with 5-fold cross-validation, Grad-CAM attention analysis, and controlled class-imbalance experiments.}
  % results
   {ConvNeXT ($M_1$) and RGC ($M_2$) form a top tier at macro-$F_1$ $0.80\pm0.02$ and $0.79\pm0.02$ respectively, a difference within one standard deviation. $M_2$ is the only model whose Grad-CAM contours consistently trace the morphological structure of RAGNs---lobes, jets, and bends---rather than defaulting to compact blobs or diffuse patterns.}
  % conclusions
   {The four-class scheme introduced here enables WAT/NAT-resolved catalogues that can serve as environment probes and progenitor classifications for diffuse cluster radio emission. The complementary strengths of $M_1$ and $M_2$---in cross-type and within-type discrimination respectively---suggest that an ensemble approach may offer a practical framework for survey-scale morphological catalogues.}


     \maketitle
%
%-------------------------------------------------------------------

\section{Introduction}\label{sec:intro}

The volume of radio continuum data is growing by orders of magnitude: from the $\sim$28\,000 sources in the original Faint Images of the Radio Sky at Twenty-Centimeters (FIRST) catalogue \citep{becker1995} to over 800\,000 in its final release \citep{proctor2011}, 4.4 million detections in the LOFAR (LOw Frequency ARray) Two-metre Sky Survey (LoTSS) \citep{shimwell2022}, and over 200\,000 sources in the Evolutionary Map of the Universe (EMU) pilot survey alone \citep{norris2021}. The trend is accelerating---EMU and LOFAR are each expected to approach 100 million sources, and the Square Kilometre Array (SKA) is projected to catalogue billions across all redshifts \citep{norris2015}, with the first SKA-Low and SKA-Mid hardware already installed in 2024. Meanwhile, multi-wavelength surveys at X-ray, optical, and infrared wavelengths are producing catalogues of comparable scale, and cross-matching radio sources with their host galaxies---once feasible by visual inspection---now demands automated methods, as demonstrated by the citizen-science Radio Galaxy Zoo (RGZ) project \citep{banfield2015,Wong2025}. Morphological classification of radio active galactic nuclei (AGNs) at scale is therefore no longer optional but a prerequisite for extracting science from current and forthcoming surveys.

In this context, artificial intelligence (AI) and deep learning (DL) in particular are necessary for classifying astronomical sources into scientifically meaningful categories at a scale that precludes visual inspection. \citet{ndung'u2023} review 32 papers published between 2017 and 2023 on the morphological classification of radio AGNs (RAGNs) using machine learning (ML), finding that convolutional neural networks (CNNs) dominate the field. They distinguish model-centric approaches---which apply novel architectures to existing, often small and low-resolution datasets---from data-centric approaches that leverage transfer learning and semi-supervised methods to exploit the larger volumes of data now available. The application of DL to RAGN classification has advanced rapidly along both fronts. Early work by \citet{Aniyan2017} demonstrated that CNNs could classify Fanaroff--Riley type I and II (FRI/FRII) and bent-tailed radio galaxies---the latter comprising wide-angle tail (WAT) and narrow-angle tail (NAT) sources---with accuracies comparable to manual inspection at a fraction of the time. Subsequent studies extended this to multi-class problems \citep[e.g.][]{Lukic2018} and to simultaneous detection and classification of multiple components within radio sources \citep[e.g.][]{Wu2019}.

Beyond purely supervised pipelines, a range of approaches have been explored. \citet{Galvin2020}, for example, used rotation- and flip-invariant self-organising maps to associate radio components without labelled training data, while \citet{Scaife2021} employed group-equivariant CNNs to reduce the dependence of model confidence on source orientation. \citet{Slijepcevic2022} tackled dataset shift through semi-supervised learning, leveraging large pools of unlabelled data to maintain classification accuracy. More recently, self-supervised and foundation models have gained traction: \citet{Slijepcevic2024} developed a multipurpose foundation model for radio astronomy that compresses unlabelled data into structured representations boosting downstream tasks in label-scarce regimes, and \citet{BaronPerez2025} expanded this paradigm by classifying radio sources into 12 morphological classes, demonstrating that learned representations capture structural diversity well beyond the standard FRI/FRII dichotomy. \citet{Lao2025} employed mask transfiners to search the FIRST survey for bent-tail radio galaxies, categorising sources into five classes that explicitly distinguish bent morphologies from straight FRI (sFRI) and straight FRII (sFRII) sources.

The present work builds on two earlier studies. \citet{hossain23} (hereafter H23) introduced RGC 0.1, a binary classifier for FRI and FRII sources combining Bootstrap Your Own Latent (BYOL) with rotation-equivariant steerable CNNs (E2CNN), following the application of BYOL by \citet{Slijepcevic2022a}. \citet{Hossain2025} (hereafter H25) constructed a curated dataset of WAT and NAT sources designed for ML applications, building on the catalogue of \citet{Sasmal2022}, and benchmarked it on five supervised models. In this paper, we scale the effort substantially: we modify the architecture of RGC 0.1, combine data from H25 with two additional catalogues of straight and bent RAGNs, and construct a completely new labelled dataset suited to simultaneous four-class classification of sFRI, sFRII, WAT, and NAT morphologies. Beyond tackling these four classes simultaneously for the first time, we show through Grad-CAM attention analysis that a compact semi-supervised model pretrained on unlabelled radio sources attends to the physical structure of RAGNs---lobes, jets, and bending---more faithfully than the largest available supervised architectures pretrained on ImageNet. We refer to the semi-supervised model as RGC 1.0 and the dataset FIRST-2060. RGC is named in honour of Radha Gobinda Chandra (1878--1975), a Bangladeshi-Indian amateur astronomer who contributed more than fifty thousand observations to the American Association of Variable Star Observers \citep{maitra2021,kapoor2023}.

The paper is organised as follows. Section~\ref{sec:ragn} reviews the morphology of RAGNs. Section~\ref{sec:data} describes the construction and preprocessing of the datasets. Section~\ref{sec:models} presents the model architectures, focusing on RGC. Section~\ref{sec:performance} reports the classification results in terms of per-class and aggregate scoring metrics, discriminative ability, and calibration. Section~\ref{sec:discussion} provides a critical discussion, covering attention map analysis, sensitivity to class imbalance, and astrophysical implications. We summarise our major findings in Section~\ref{sec:conclusions}.

\begin{table*}
\caption{Major surveys from array radio telescopes (ART) that have catalogued significant numbers of radio-loud AGNs (RAGNs).}
\label{tab:ragns}
\centering
\begin{tabular}{l l c c r r r r r l}
\hline\hline
ART & Survey & Sensitivity & Resolution & RAGN & FRI & FRII & WAT & NAT & Reference \\
\hline
VLA     & FIRST   & 150  & $5''$ (1400)  & 717  &      &      & 430  & 287 & \citet{Sasmal2022} \\
        &         &      &               & 4876 &      &      & 4424 & 652 & \citet{Lao2025} \\
        &         &      &               & 1222 & 591  & 631  &      &     & \citet{Porter2023} \\
LOFAR   & LoTSS   & 83   &  $6''$ (144)  & 5805 & 1256 & 423  &      &     & \citet{Mingo2019} \\
        &         &      &               & 55   &      &      & 45   & 10  & \citet{Pal2023} \\
ATCA    & ATLAS   & 20   & $12''$ (1400) & 45   &      &      &      &     & \citet{Dehghan2014} \\
GMRT    & TGSS    & 3500 & $25''$ (150)  & 264  &      &      & 203  & 61  & \citet{bhukta2022} \\
MeerKAT & MIGHTEE & 6    & $5''$ (1250)  & 359  &      &      &   &  & \citet{Vardoulaki2025} \\
\hline
\end{tabular}
\tablefoot{Sensitivity is given as `median rms' in $\mu$Jy beam$^{-1}$. Resolution is in arcsec, with the frequency given within brackets in MHz.}
\end{table*}

\section{Radio active galactic nuclei classification} \label{sec:ragn}

AGNs are powered by accretion onto supermassive black holes (SMBHs), with their luminosity driven by the release of gravitational potential energy from infalling material. Magnetic fields can additionally extract the rotational energy of a spinning SMBH via the Blandford--Znajek process, powering relativistic jets along the polar axes \citep{Blandford1977}. These bipolar jets and their associated lobes emit radio-frequency radiation via the synchrotron mechanism, characterising radio-loud AGNs, or RAGNs \citep{Urry1995}. The central engine occupies less than one light-year (ly), host galaxies typically span $\sim$100\,kly, and the radio jets can extend to several Mly.

RAGNs are traditionally classified by their surface brightness distribution: FRI sources exhibit brightness decreasing from the centre toward the edges, while FRII sources show the opposite, often terminating in distinct hotspots \citep{Fanaroff1974}. This FRI/FRII dichotomy correlates with jet kinetic power, accretion state, and environment, with FRI sources generally associated with lower-power jets and denser environments \citep{Hardcastle2007}. Hybrid FRI-FRII morphologies also exist, as do FR0 sources, which feature a bright core but lack clearly detectable extended emission. In some instances, particularly among FRI sources, the radio jets appear deflected or bent from their initial trajectory. If the opening angle $\theta_\mathrm{OA}$ between the two tails is less than $\sim90^\circ$, forming a characteristic `V' shape, the source is classified as a narrow-angle-tail \citep[NAT;][]{rudnick1976}; if $\theta_\mathrm{OA} \gtrsim 90^\circ$, forming a `C' shape, it is termed a wide-angle-tail \citep[WAT;][]{Owen1976, ODea2023}. The precise classification is sensitive to projection effects \citep{proctor2011} and to the sensitivity and angular resolution of the observing telescope.

Some RAGNs display a transition from collimated jets to diffuse tails, likely marking the point where the jet leaves the interstellar medium (ISM) of the host galaxy and enters the intracluster medium (ICM) or intergalactic medium (IGM) \citep{ODea2023}. Understanding such distortions is a primary motivator for RAGN research, as bent jets serve as anemometers for the dense environments of galaxy groups and clusters. \citet{Vardoulaki2025} attribute jet distortions to several mechanisms, including interactions with the IGM \citep{miley1972}, buoyancy forces, jet precession, gravitational interactions with neighbouring galaxies, and passage through steep pressure gradients. Ram pressure in the IGM and ICM is particularly well studied as a driver of tail bending. \citet{Garon2019}, \citet{Mingo2019}, and \citet{Golden-Marx2021} demonstrate that bent RAGNs are found predominantly in dense environments up to a redshift of $z \approx 2.2$, and that their prevalence and degree of bending both increase toward the centres of massive, rich clusters \citep{Vardoulaki2025}.

A substantial number of bent RAGNs have been found in surveys conducted by different array radio telescopes: FIRST by the Karl G. Jansky Very Large Array (VLA), LoTSS by LOFAR, the Australia Telescope Large Area Survey (ATLAS) by the Australia Telescope Compact Array (ATCA), the TIFR GMRT Sky Survey (TGSS), and the MeerKAT International GHz Tiered Extragalactic Explorations (MIGHTEE). The source counts from each survey and the corresponding references are given in Table~\ref{tab:ragns}, which shows that FIRST has produced by far the largest number of catalogued bent RAGNs.

Consequently, the FIRST survey has been the focus of numerous investigations. \citet{proctor2011} visually inspected over 800,000 sources, identifying 7,106 multi-component candidates and ultimately classifying approximately 400 as WATs or NATs. \citet{Miraghaei2017} constructed a batched dataset of 1,256 FRI and FRII sources, later extended by \citet{Porter2023}, which has since become a community standard for ML studies of RAGNs. A categorised WAT and NAT sample was produced by \citet{Sasmal2022}, who filtered the FIRST 2014 data release for sources with angular sizes greater than $10''$; from approximately 95,000 candidates, visual inspection of tail opening angles yielded 717 classified sources. H25 subsequently refined this catalogue and prepared a batched dataset of 639 WATs and NATs for use in ML. \citet{Lao2025} presented a much larger dataset of approximately 4,800 bent RAGNs from FIRST, though heavily skewed toward WATs, with only 13\% classified as NATs. They trained the Radio Galaxies Classification with Mask Transfiner (RGCMT) on 3,600 manually annotated sources and used it to classify nearly one million FIRST sources, producing 11,000 machine-identified candidates that were then verified by visual inspection to yield the final catalogue of 4,876 bent RAGNs.

Visually inspected datasets are now widely available for binary classification of RAGNs into FRI and FRII, or WAT and NAT. Some works have incorporated FR0, compact, and/or bent sources into multiclass schemes \citep[e.g.][]{Maslej2021} or fused separate binary classifiers \citep[e.g.][]{Aniyan2017}, while others defined classes by the number of radio components and peaks rather than by physical morphology \citep[e.g.][]{Wu2019}. However, to our knowledge, no published multiclass model classifies RAGNs simultaneously into sFRI, sFRII, WAT, and NAT as four morphologically distinct classes: existing works treat WAT and NAT as a single undifferentiated `Bent' class alongside FRI, FRII, and Compact. Because we attempt this four-class separation here for the first time, it is worth examining the rationale behind choosing these specific classes.

FRI and FRII sources can exhibit both straight and bent morphologies, so the classes FRI, FRII, WAT, and NAT are not mutually exclusive. Classifying sources into FRI, FRII, and Bent therefore requires either accepting overlapping cases or first separating straight from bent sources within the FR classes. Our scheme avoids this ambiguity at the level of morphological labelling by restricting the FR classes to straight sources only, designating them sFRI and sFRII---similar in spirit to \citet{Lao2025}, but differing in that we resolve the Bent class into WAT and NAT rather than treating it as a single category. We did not include compact, FR0, or single-tailed sources. Single-tailed sources are excluded by data scarcity: sFRI, sFRII, WAT, and NAT were the only classes for which we could assemble reliable samples of comparable size through multi-person visual inspection. The exclusion of compact and FR0 sources is more physical. FR0 sources are genuinely compact, while an intrinsically extended source may appear compact at sufficiently high redshift. Distinguishing these two cases requires redshift information, making compact source classification inherently multimodal---beyond the scope of the image-only framework presented here.

\begin{figure*}
    \centering
    \includegraphics[width = \linewidth]{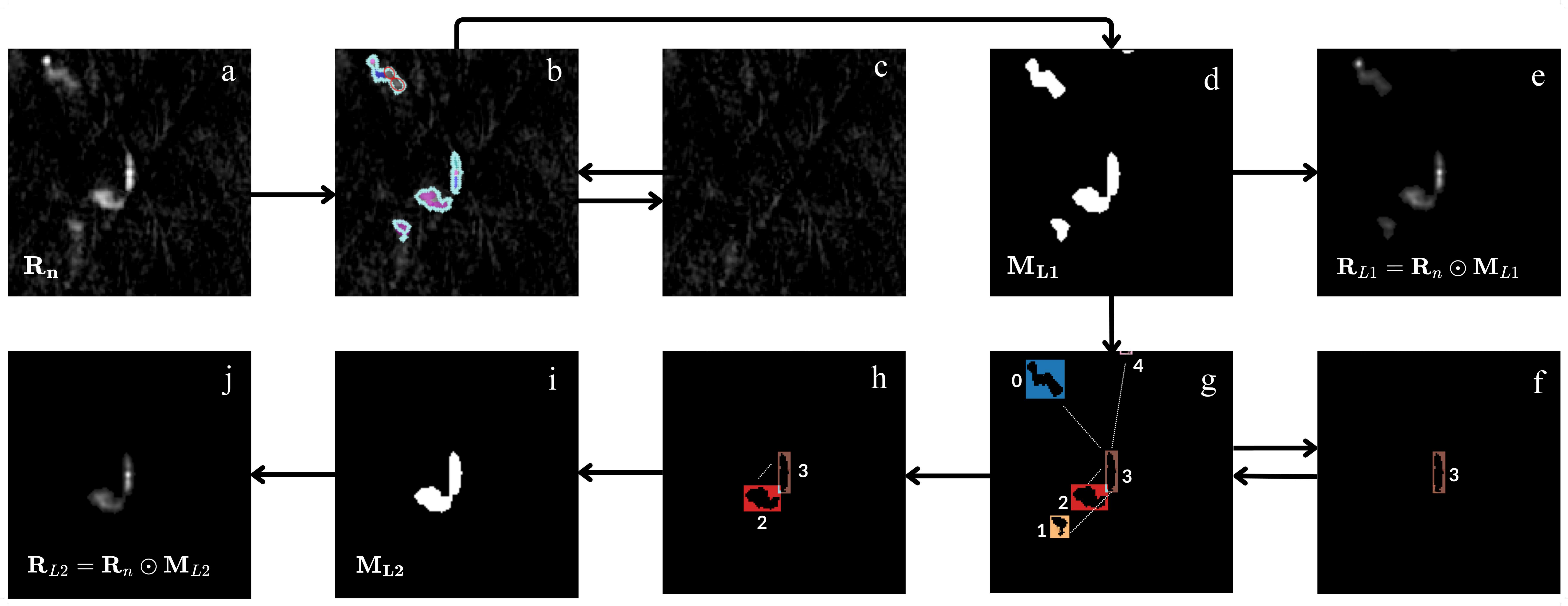}
    \caption{The steps in our \texttt{PyBDSF} preprocessing pipeline as described in Section~\ref{sec:prep}: (a) original normalised image $\mathbf{R}_n$, (b) identified islands and sources, (c) residual image after subtracting the identified islands from (a), (d) binary mask associated with the identified islands, (e) masked image $\mathbf{R}_{L1}$ retaining all islands, (f) the central island centred on the target source, (g) distance map of all islands relative to the central island, (h) islands identified as neighbours of the central island, (i) refined mask after removing spurious islands, (j) masked image $\mathbf{R}_{L2}$ with spurious sources removed.}
    \label{fig:pybdsf}
\end{figure*}

\section{Data and preprocessing} \label{sec:data}
Both our labelled ($\mathbf{R}_L$) and unlabelled ($\mathbf{R}_U$) datasets are constructed from FIRST survey images; $\mathbf{R}_L$ is used for supervised training and fine-tuning and $\mathbf{R}_U$ for self-supervised pre-training. FIRST provides an angular resolution of $5$ arcsec at 1400 MHz (Table~\ref{tab:ragns}), and the downloaded image products have a native pixel scale of $1.8$ arcsec pixel$^{-1}$. For training and testing, we employ image cutouts of $150 \times 150$ pixels, corresponding to an angular extent of $4.5$ arcmin on a side. This size represents a compromise between field of view and target scale: larger cutouts would render the target comparatively small within the image, while smaller cutouts would risk truncating extended emission or failing to fully encompass some sources. The preparation of each dataset is described below.

\subsection{Unlabelled data}

For self-supervised pre-training, we employ the batched dataset of 20,000 radio sources constructed for ML applications by \citet{Slijepcevic2022} from the Radio Galaxy Zoo \citep[RGZ;][]{Wong2025}. RGZ Data Release 1 (DR1) comprises approximately 100,000 sources, of which 99.2\% originate from FIRST. For each FIRST source with a largest angular size ($\theta_{LAS}$) between 15 and 270 arcsec, \citet{Slijepcevic2022} retrieved the corresponding radio image via the \texttt{Python} API of \texttt{SkyView},\footnote{\url{https://skyview.gsfc.nasa.gov}.} a virtual observatory service providing access to astronomical imaging data from multiple surveys. The original $300^2$-pixel images were cropped to retain only the central $150^2$ pixels, and all pixels at radii greater than $0.6\,\theta_{LAS}$ were set to zero. A $3\sigma$ amplitude threshold was then applied, and each image $\mathbf{R}$ was normalised to $\mathbf{R}_n$ for storage in PNG format.
Unresolved sources were removed by enforcing a hard lower threshold of $\theta_{\rm cut}=28''$ on source angular extension; this value is the smallest cut that minimises the Fr\'echet distance between feature-space representations of the unlabelled RGZ DR1 and the labelled MiraBest dataset \citep{Porter2023}, also employed in H23.

The resulting 20,000 sources are partitioned into 10 training batches and one test batch. We retrieved these data directly from the corresponding GitHub repository\footnote{\url{https://github.com/inigoval/fixmatch}} without any additional processing; hereafter this dataset is denoted $\mathbf{R}_U$. We note that \citet{Slijepcevic2024} employed a considerably larger unlabelled set of 108,000 RGZ sources for self-supervised learning; pre-training at that scale is beyond the computational budget of the present study.

\subsection{Labelled data}

Unlike $\mathbf{R}_U$, our labelled dataset $\mathbf{R}_L$ is constructed here for the first time by combining selected sources from three catalogues: MiraBest \citep{Porter2023}, the Sasmal catalogue \citep{Sasmal2022}, and the Lao catalogue \citep{Lao2025}, as shown in Table~\ref{tab:ragns}. These catalogues contain sources spanning the FRI, FRII, WAT, and NAT classes, from which we visually selected reliable samples of sFRI, sFRII, WAT, and NAT and preprocessed the corresponding FIRST images for use in ML.

\subsubsection{Visual inspection} \label{sec:vis}

Each catalogue was accessed from \texttt{VizieR} using the \texttt{astroquery} module of \texttt{astropy} and converted to a \texttt{Pandas DataFrame} for further processing. The source coordinates were then used to retrieve the corresponding FIRST images in \texttt{FITS} (Flexible Image Transport System) format via \texttt{SkyView} within \texttt{astroquery}. To facilitate visual inspection, we compiled three PDF files---one per catalogue---in which each page displayed 30 images arranged in a $5\times 6$ grid, with the serial number and original catalogue label shown in the top-left corner of each image.

Six co-authors carried out the visual selection in a group, with four individuals each assigned to one of the four target classes. Each person reviewed all relevant PDF files and identified the most suitable candidates within their assigned class. Classification was based primarily on the opening angle $\theta_{OA}$: sources with $\theta_{OA} \approx 180^\circ$ were labelled straight, those with $90^\circ \lessapprox \theta_{OA} \lessapprox 180^\circ$ were labelled WAT, and those with $\theta_{OA} \lessapprox 90^\circ$ were labelled NAT. Straight sources were further divided into sFRI and sFRII based on the brightness distribution of their tails, as described in Section~\ref{sec:ragn}. After this initial selection, the class assignments were reshuffled and each individual's selections were independently reviewed by another member of the group. Following this two-tier process, four PDF files were compiled---one per class---and the remaining two members of the group, including one astronomer, jointly reviewed all four files, critically evaluating each classification decision. The images that survived this final review were passed on to the preprocessing step.

\subsubsection{Preprocessing} \label{sec:prep}

\begin{table}
\caption{Distribution of the four RAGN classes across the three source catalogues that together constitute our labelled datasets.}
\label{tab:data}
\centering
\begin{tabular}{lcccc c}
\hline\hline
Catalogue & sFRI & sFRII & WAT & NAT & Catalogue total \\
\hline\hline
Lao      & 96  & 168 & 436 & 234 & 934 \\
MiraBest & 224 & 420 & 0   & 0   & 644 \\ 
Sasmal   & 2   & 0   & 239 & 241 & 482 \\
\hline\hline
\textbf{Total} & 322 & 588 & 675 & 475 & 2060 \\
\hline\hline
\end{tabular}
\end{table}

\begin{figure*}
    \centering
    \includegraphics[width=\textwidth]{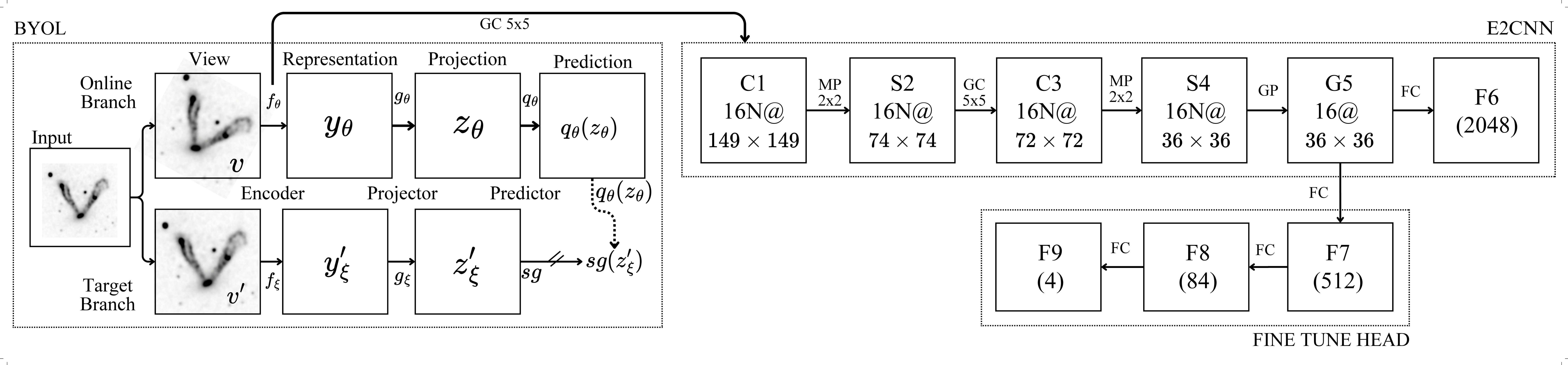}
    \caption{Architecture of the RGC model. \textit{Left}: The BYOL self-supervised pre-training architecture, with E2CNN as the encoder $f_\theta$. \textit{Right}: The E2CNN encoder followed by the fine-tune head. GC, MP, GP, and FC denote group convolution, max-pooling, group-pooling, and fully connected, respectively.}
    \label{fig:rgc-model}
\end{figure*}

Because of the way the images were downloaded, the target source is always positioned at the centre. We preprocessed all images using \texttt{PyBDSF} (the Python Blob Detector and Source Finder; \citealt{Mohan2015}) to produce two sets of processed images: $\mathbf{R}_{L1}$, which retains spurious sources, and $\mathbf{R}_{L2}$, from which spurious sources have been removed. The main steps of this pipeline are illustrated in Fig.~\ref{fig:pybdsf} and described below.

The method of creating $\mathbf{R}_{L1}$ using \texttt{PyBDSF} follows that of H25, with one addition: here we include a wavelet decomposition using the \`{a} trous algorithm. As illustrated in Fig.~\ref{fig:pybdsf}, the pipeline proceeds as follows.
The background of each image is first estimated, and pixels exceeding a $5\sigma$ peak threshold are grouped into distinct islands of candidate sources traced down to a $3\sigma$ boundary, as shown in panel (b). Source detection and fitting are performed across multiple wavelet scales. The number of scales is set automatically per image using PyBDSF's default à trous configuration, which derives the maximum wavelet order from the size of the largest island; each island is then fitted by a two-dimensional Gaussian at every scale.
If the resulting residual image (c) is free of residual sources and peaks, the islands are accepted and a binary mask is constructed from their extent (d). Multiplying this mask by the original normalised image (a) yields the final masked image (e), which constitutes the $\mathbf{R}_{L1}$ dataset.

For the second dataset $\mathbf{R}_{L2}$, which was not present in H25, we resume from the binary mask in panel (d) and assign a unique serial number to each identified island, as shown in panel (g). The central island---corresponding to the central part of the target source---is then isolated (f), and the distance of every other island from its centre is measured (g). Islands lying within approximately 30 pixels of the central island are retained as neighbours (h), while all remaining islands are discarded as spurious. 
At the FIRST pixel scale of $1.8''$, this corresponds to $\sim 54''$, comparable to the typical angular extent of most RAGN structures in our sample.
This threshold worked well for the majority of sources, as verified by visual inspection, but required manual adjustment for fewer than 10\% of the images. The result is the refined binary mask $\mathbf{M}_{L2}$ shown in panel (i), which when multiplied by the original normalised image (a) produces the final masked image without spurious sources (j), constituting the second dataset $\mathbf{R}_{L2}$.

All FITS files processed by the automated pipeline were compiled into four PDF files corresponding to the four classes, following the same format as the PDF files used during the initial visual inspection. These files were reviewed by the same two individuals who had conducted the final review of the original images. This second review was necessary for two reasons: the preprocessing occasionally crops the target source or fails to discard all spurious sources, and the morphological characteristics of each source become considerably clearer on the masked images than on the originals. During this review, sources were flagged for one of three outcomes: discarding due to poor preprocessing, discarding due to misclassification, or reassignment to a different class due to ambiguity in the initial labelling. This process yielded our final labelled datasets, converted to PNG format, and the number of sources in each of the four classes drawn from each of the three catalogues is listed in Table~\ref{tab:data}.

It shows that the Lao catalogue is the dominant contributor, followed by MiraBest and then the Sasmal catalogue. Of the more than 600 WATs and NATs released by H25, we retain 482 in this refined dataset. Although the Lao catalogue contains over 4,000 WATs, we could not retain a proportionate number for two reasons: closer visual inspection reclassified some WATs as straight or NAT sources, and the total number of WATs needed to remain reasonably comparable to the sample sizes of the other classes. For each class, we included only the most unambiguous cases. Nevertheless, one should bear in mind that opening angles vary continuously from $0^\circ$ to $180^\circ$---giving rise to hybrid WAT-NAT morphologies---and that the brightness distribution of tails likewise forms a continuum, making the sFRI-sFRII boundary inherently gradual. Our final dataset contains 2060 sources in total, with WATs being the most numerous and sFRI the least. A catalogue of these sources is available in our GitHub repository, and the preprocessed, ML-ready images derived from this catalogue are referred to hereafter as FIRST-2060, containing both $\mathbf{R}_{L1}$ and $\mathbf{R}_{L2}$.

\begin{table*}
\centering
\small
\caption{Architectural properties and training hyperparameters of the six models benchmarked in this work.}
\label{tab:models}
\begin{tabular}{llcccccccc}
\hline\hline
ID & Model & Depth & Params & Optimizer & Learning rate & Batch size & Weight decay & Scheduler & Epochs \\
\hline
$M_1$ & ConvNeXT-B  & 28 & 89  & Adam  & $10^{-4}$ & 32 & --- & StepLR & 30 \\

$M_2$ & RGC (pretraining)  & 7  & 42  & Adam  & $3\times10^{-4}$ & 16 & --- & --- & 500 \\
      & RGC (finetuning)   & 5  & 11  & Adam  & $8.5\times10^{-4}$ & 8 & $2.12\times10^{-6}$ & CALR & 1000 \\

$M_3$ & VGG-16    & 16 & 138 & Adam  & $10^{-4}$ & 32 & --- & StepLR & 30 \\

$M_4$ & Swin-B    & 12 & 88  & AdamW & $10^{-4}$ & 32 & --- & CALR & 30 \\

$M_5$ & ResNet-50 & 50 & 25  & Adam  & $10^{-4}$ & 32 & --- & StepLR & 30 \\

$M_6$ & ViT-B/16  & 12 & 86  & AdamW & $10^{-4}$ & 32 & --- & CALR & 30 \\
\hline\hline
\end{tabular}
\tablefoot{Depth refers to the number of weight layers for CNN-based models and transformer blocks for attention-based models. Parameter counts are in millions and are approximate. CALR: CosineAnnealingLR.}
\end{table*}

\section{The classification models}  \label{sec:models}

\subsection{The RGC model}

The RGC model combines different existing DL architectures into a framework for classifying RAGNs. Its initial version (RGC 0.1), described in H23, used self-supervised learning to classify RAGNs into FRI and FRII types. Here we present RGC 1.0, which extends this to the four-class problem of sFRI, sFRII, WAT, and NAT. As in H23, we combine BYOL \citep{Grill2020} with an $E(2)$-equivariant steerable CNN \citep[E2CNN;][]{Weiler2019, Scaife2021} as the encoder, since BYOL was found to outperform SimCLR. The model operates in two stages, illustrated in Fig.~\ref{fig:rgc-model}: a task-agnostic self-supervised pre-training stage on $\mathbf{R}_U$, followed by a task-specific supervised fine-tuning stage on $\mathbf{R}_{L1}$ and $\mathbf{R}_{L2}$.

In the pre-training stage, BYOL learns a representation by training an online network---comprising an encoder $f_\theta$, projector $g_\theta$, and predictor $q_\theta$---against a momentum-updated target network with weights $\xi$. Given an image $x$, two augmented views $v$ and $v'$ are produced; the online network predicts the target network's projection, and the loss is the negative cosine similarity between the two $l_2$-normalised vectors \citep{Grill2020}. The encoder $f_\theta$ is E2CNN, which preserves invariance under the Euclidean group $E(2)$. Following \citet{Scaife2021}, it comprises group convolution layers C1 and C3, subsampling layers S2 and S4, and a group-pooling layer G5, producing a feature vector of dimension 2048 at F6. In the fine-tuning stage, the final FC layer is replaced by a classification head of three fully connected layers (F7: 512, F8: 84, F9: 4), where F9 produces four logits corresponding to the four classes. Class probabilities are obtained via softmax, and the model is trained by minimising the cross-entropy loss.

\subsection{Supervised baselines}

To benchmark RGC 1.0, we compare it against five fully supervised baseline models that span the main families of modern DL architectures for image classification. All baselines are initialised with weights pretrained on ImageNet. The $150\times150$ pixel grayscale images are resized to $224\times224$ pixels and converted to three-channel inputs by duplicating the grayscale channel to meet the input requirements of each architecture.

The first two baselines are classical CNN architectures. VGG-16 \citep{Simonyan2015} is a deep network of 16 weight layers built from $3\times3$ convolutions and max-pooling, followed by fully connected layers. ResNet-50 \citep{He2016} introduced residual connections that allow gradients to flow through deeper networks, comprising 50 layers organised into four residual stages. The third baseline, ConvNeXT \citep{Liu2022}, is a modernised CNN that borrows design choices from vision transformers---such as depthwise convolutions and layer normalisation---while retaining a purely convolutional pipeline.

The remaining two baselines are transformer-based. The Vision Transformer \citep[ViT-B/16;][]{Dosovitskiy2021} divides the input image into non-overlapping $16\times16$ pixel patches and processes them with a multi-head self-attention encoder. The Swin Transformer \citep[Swin-B;][]{Liu2021} reduces the cost of full self-attention by computing it within local shifted windows and merging spatial tokens hierarchically, combining local inductive biases with long-range modelling capacity. Throughout this paper, we refer to ConvNeXT, VGG-16, Swin-B, ResNet-50, and ViT-B/16 as $M_1$, $M_3$, $M_4$, $M_5$, and $M_6$ respectively, with $M_2$ reserved for the RGC model; the full list of architectures is given in Table~\ref{tab:models}.

\begin{figure*}
    \centering
    \includegraphics[width=\textwidth]{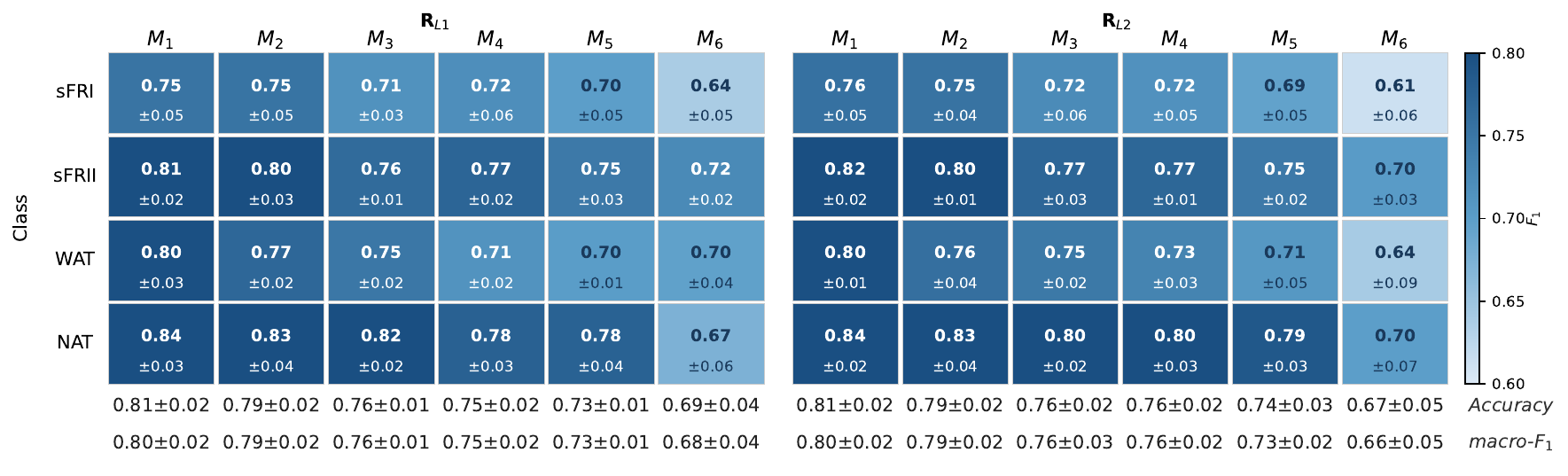}
    \caption{Per-class $F_1$-scores, accuracy, and macro-$F_1$ for all six models on $\mathbf{R}_{L1}$ (left) and $\mathbf{R}_{L2}$ (right), averaged over 5-fold cross-validation. Standard deviations are shown below each score. The colour scale maps $F_1$ values from high (dark) to low (light).}
    \label{fig:f1-heatmap}
\end{figure*}

\subsection{Training details}

We pretrained the RGC model ($M_2$) on the unlabelled dataset $\mathbf{R}_U$ for 500 epochs. The pretraining was repeated with varying hyperparameters, and the final set (the second row of Table~\ref{tab:models}) was selected based on the loss curve and validation accuracy. The learning rate was kept small to mitigate overfitting, with the batch size chosen accordingly; because learning proceeded slowly at this rate, weight decay was unnecessary.

For fine-tuning, we first determined the architecture of the classification head by varying the number and width of the fully connected layers and selecting the configuration that performed best based on visual inspection of training curves in Weights \& Biases,\footnote{\url{https://wandb.ai}} a cloud-based experiment-tracking platform that logs training metrics, hyperparameters, and model artefacts in real time. With the architecture fixed, we then used Optuna \citep{Akiba2019} to optimise the remaining hyperparameters. Optuna performs sequential model-based optimisation: given bounds and an optional step size for each hyperparameter, it proposes a trial configuration, evaluates its performance, and uses the result to inform subsequent trials. We ran approximately 700 trials to identify the final set of hyperparameters (Table~\ref{tab:models}).

For the five supervised baselines ($M_1$, $M_3$--$M_6$), we downloaded each architecture pretrained on ImageNet-1k \citep{Deng2009}---1.28 million images across 1000 classes---from \texttt{torchvision} and appended a fine-tuning classification head following the same general procedure as for $M_2$. Unlike $M_2$, extensive hyperparameter optimisation with Optuna was not necessary, as the standard values (Table~\ref{tab:models}) already yielded strong performance and further tuning produced negligible changes.
This is evidenced in Fig.~\ref{fig:loss}, where training and validation losses are plotted against epoch for all six models. Only $M_2$ exhibits sustained convergence; the supervised baselines plateau within the first few tens of epochs and show no further improvement.

All six models were evaluated using 5-fold cross-validation: the labelled dataset was divided into five folds of 412 images each, with four folds used for training and the remaining fold for testing in each run. The test fold was rotated across runs so that every image appeared in exactly one test batch by the end of the five runs, with no images shared between training and test splits in any individual run. All training and testing were performed on \texttt{Timaeus}, a Supermicro workstation equipped with an NVIDIA Leadtek Quadro RTX A4000 GPU with 16\,GB of memory.

\section{Performance of the RGC model} \label{sec:performance}

\subsection{Classification metrics}

The $F_1$-score for a given class is the harmonic mean of precision and recall, where precision is the fraction of predicted positives that are true positives and recall is the fraction of actual positives that are correctly identified; the harmonic mean ensures that both quantities must be high for $F_1$ to be high, penalising models that sacrifice one for the other. Macro-$F_1$ averages the per-class $F_1$-scores with equal weight, treating minority and majority classes alike, while accuracy is simply the fraction of all samples classified correctly. Fig.~\ref{fig:f1-heatmap} shows the per-class $F_1$-scores, accuracy, and macro-$F_1$ for all six models on both $\mathbf{R}_{L1}$ and $\mathbf{R}_{L2}$, averaged over 5-fold cross-validation with standard deviations reported. With uncertainties of $\pm0.01$ to $\pm0.06$ across most cells, the ranking is stable, though the models naturally divide into two tiers: $M_1$ and $M_2$ form a top cluster whose internal difference ($0.80$ vs.\ $0.79$) is within one standard deviation, and $M_3$--$M_6$ form a second cluster with similarly small internal separations (except $M_6$), while the gap between the two tiers---$M_2$ at $0.79$ versus $M_3$ at $0.76$ (macro-$F_1$)---exceeds the uncertainty of either model. The overall ranking is consistent across both dataset regimes: $M_1 > M_2 > M_3 > M_4 > M_5 > M_6$ in macro-$F_1$ and accuracy.

The RGC model $M_2$ ranks second overall with a macro-$F_1$ of $0.79\pm0.02$, just $0.01$ behind $M_1$ ($0.80\pm0.02$)---a difference smaller than one standard deviation of either model. This near-parity is noteworthy for three reasons. First, $M_2$ has 53 million parameters compared to 89 million for $M_1$, and the majority of those (42 million) are used during BYOL pre-training. Second, $M_2$ was pre-trained on unlabelled astronomical images rather than the large labelled ImageNet dataset used to initialise all five baselines. Third, $M_2$ incorporates rotational equivariance as a physical prior, encoding the fact that the morphological class of a radio source is independent of its orientation on the sky. The per-class $F_1$-scores of $M_2$ reflect this competitiveness: on $\mathbf{R}_{L1}$, it achieves $0.75\pm0.05$ (sFRI), $0.80\pm0.03$ (sFRII), $0.77\pm0.02$ (WAT), and $0.83\pm0.04$ (NAT), none of which differ from $M_1$ by more than one standard deviation of either model.

The most consistent pattern across all models is the class-dependent difficulty of the task. NAT achieves the highest per-class $F_1$-score for every model on both datasets---reaching $\sim 0.84\pm0.02$ for $M_1$ on $\mathbf{R}_{L1}$---while sFRI is the hardest class, with $F_1$-scores ranging from $0.61\pm0.06$ ($M_6$) to $0.76\pm0.05$ ($M_1$, $\mathbf{R}_{L2}$). This difficulty ordering is expected given morphological continuity between FRI and FRII brightness distributions: the confusion matrices (Fig.~\ref{fig:conf1} in Appendix \ref{app:conf}) confirm that sFRI sources are most frequently confused with sFRII and vice versa (in terms of fraction of sources misclassified). WAT and sFRII occupy intermediate positions in difficulty, consistent across models and datasets. The transformer-based models $M_4$ and $M_6$ show the largest standard deviations in WAT $F_1$-scores ($\pm0.02$--$\pm0.09$), suggesting greater fold-to-fold variance.

Comparing the two panels of Fig.~\ref{fig:f1-heatmap} ($\mathbf{R}_{L1}$ and $\mathbf{R}_{L2}$), $M_1$ improves in two classes (sFRI, sFRII, within one standard deviation) and is unchanged in the other two. $M_2$ drops negligibly in one class and is unchanged in the other three---the most stable behaviour of any model. $M_3$ improves in two classes (sFRI, sFRII) and drops in one. $M_4$ and $M_5$ show at most one class changing by 0.01 in either direction. $M_6$, by contrast, drops in two classes and improves in one, producing the only meaningful macro-$F_1$ decline ($0.68\to0.66$) and a near-doubling of the WAT standard deviation ($\pm0.04\to\pm0.09$). Overall, the per-class profiles are nearly unchanged across regimes for all models, but most notably for $M_1$ and $M_2$, whose macro-$F_1$ scores are effectively invariant to the presence or absence of spurious sources.

The confusion matrices (Fig.~\ref{fig:conf1}) add one further detail: for $M_6$, the WAT diagonal drops from 103 to 88 between $\mathbf{R}_{L1}$ and $\mathbf{R}_{L2}$---a loss of 15 correctly classified WAT sources---while the NAT diagonal improves from 55 to 62, suggesting that $M_6$ redistributes predictions from WAT to NAT once spurious sources are removed rather than improving uniformly. This trade-off is largely absent in $M_1$ and $M_2$, whose WAT and NAT diagonals remain stable at 111/80 and 110/79 on $\mathbf{R}_{L1}$ versus 111/80 and 106/81 on $\mathbf{R}_{L2}$; the 4-source WAT drop for $M_2$ is modest compared to the 15-source loss for $M_6$, confirming that the regime-insensitivity of these two models is genuine.

\begin{figure*}
  \centering
  \includegraphics[width=0.49\linewidth]{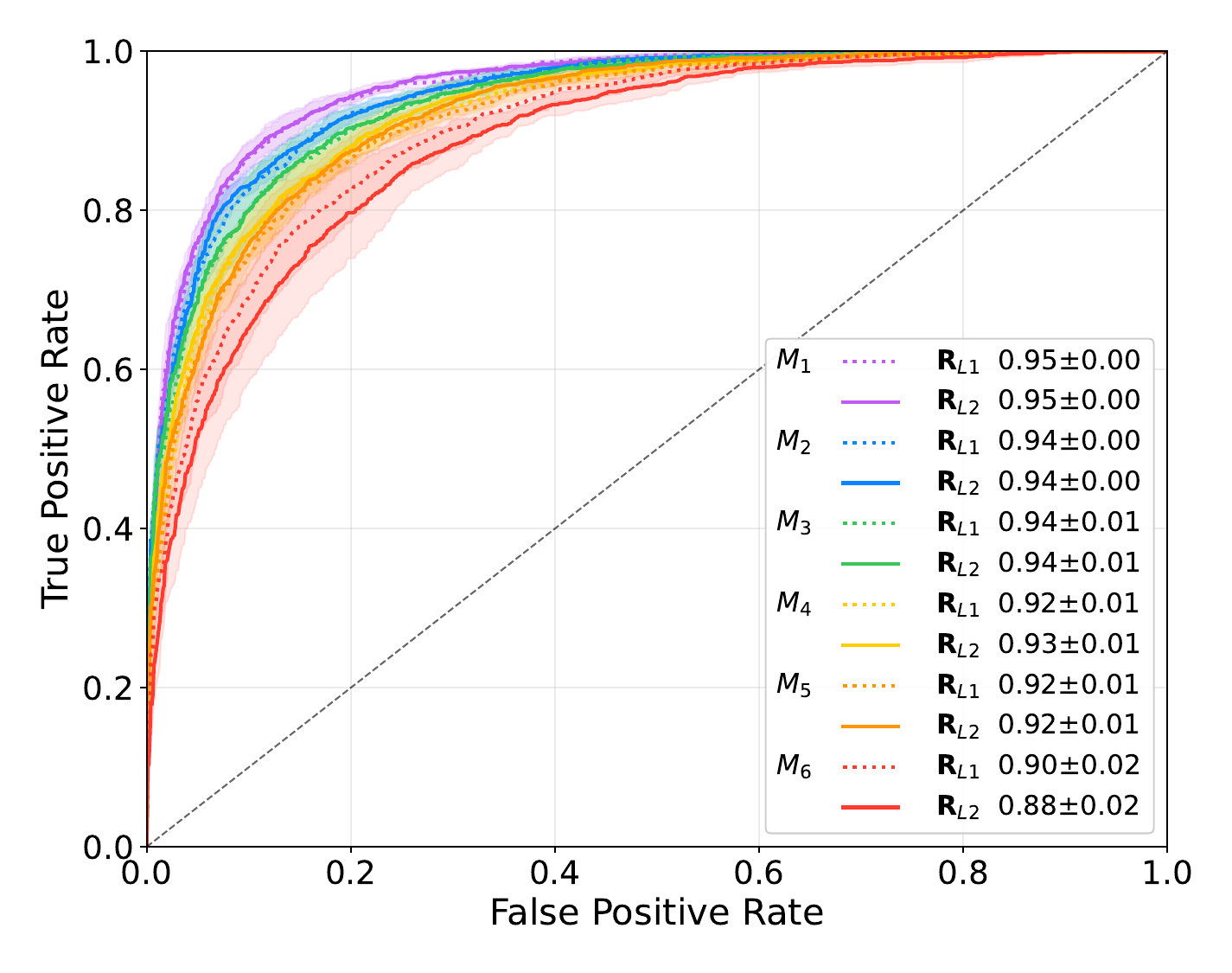}\hfill
  \includegraphics[width=0.49\linewidth]{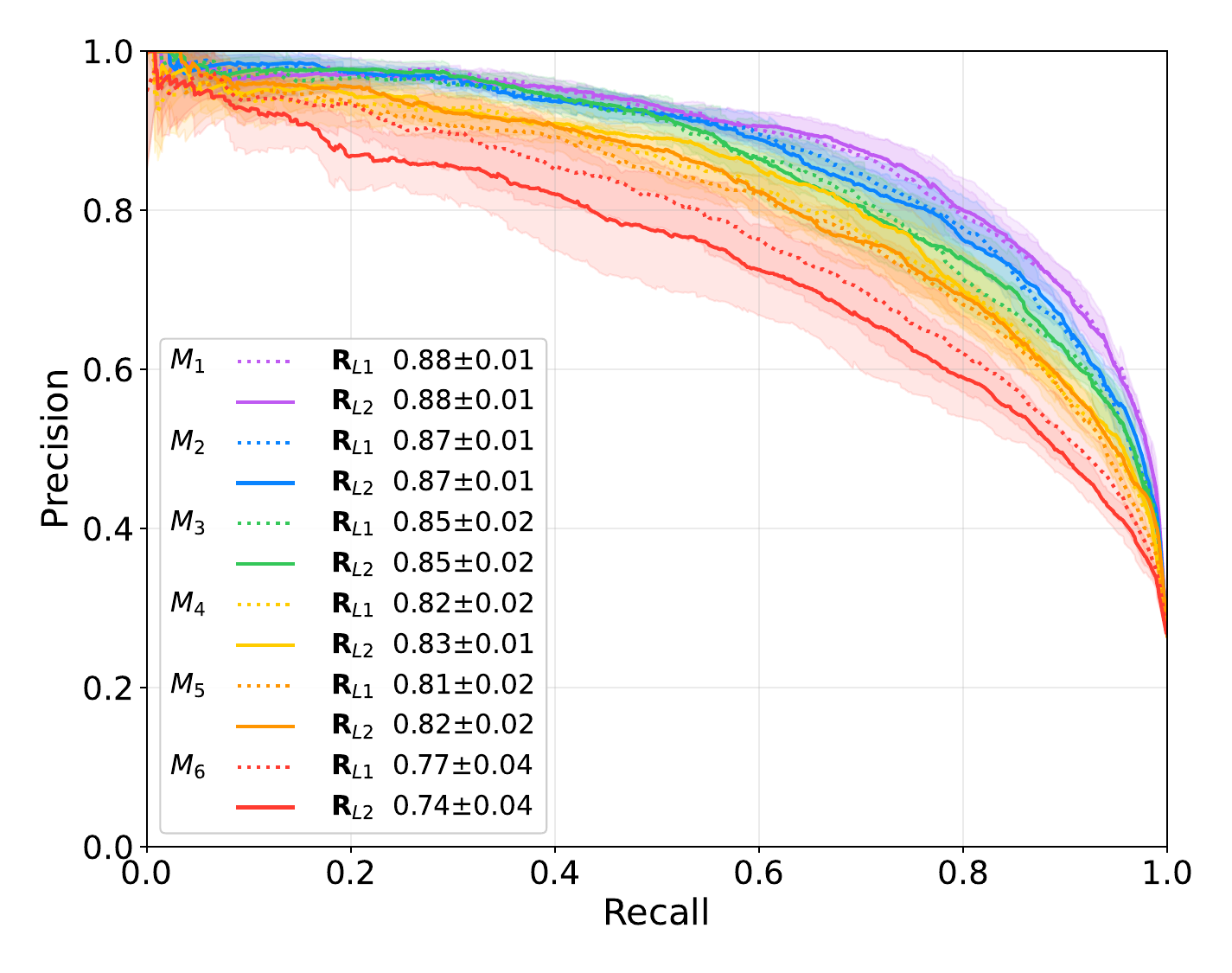}
  \caption{\textit{Left}: ROC curves and AUC-ROC scores for all six models on $\mathbf{R}_{L1}$ (dotted) and $\mathbf{R}_{L2}$ (solid), averaged over 5-fold cross-validation; shaded bands show $\pm1$ standard deviation. \textit{Right}: Precision-recall curves and AUC-PR scores with the same conventions.}
  \label{fig:roc-pr}
\end{figure*}

\begin{figure}
    \centering
    \includegraphics[width=\columnwidth]{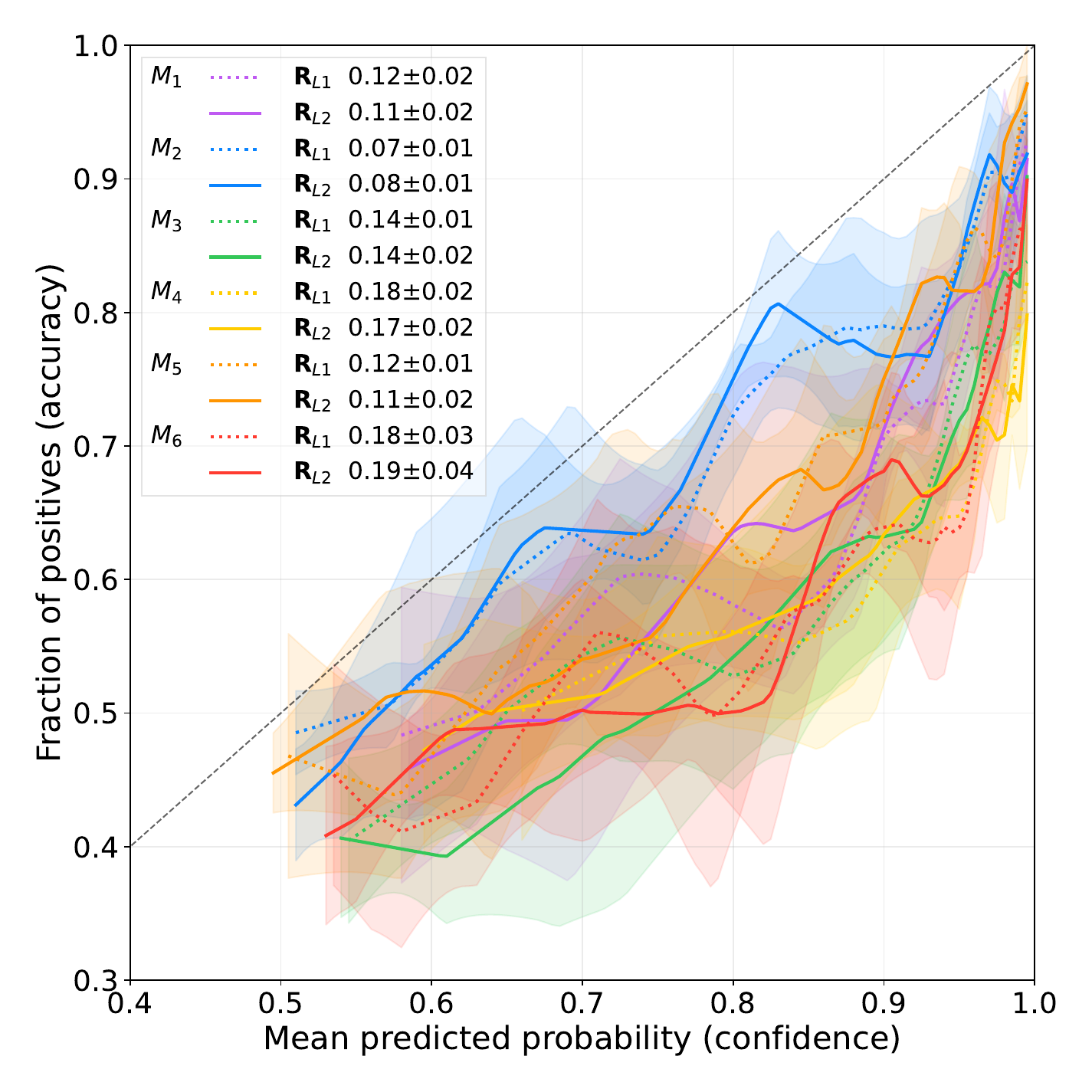}
\caption{Reliability diagrams and ECE for all six models on $\mathbf{R}_{L1}$ (dotted) and $\mathbf{R}_{L2}$ (solid), averaged over 5-fold cross-validation. Colours, line styles, and shaded bands follow Fig.~\ref{fig:roc-pr}.}
    \label{fig:ece}
\end{figure}

\begin{figure*}
    \centering
    \includegraphics[width=\textwidth]{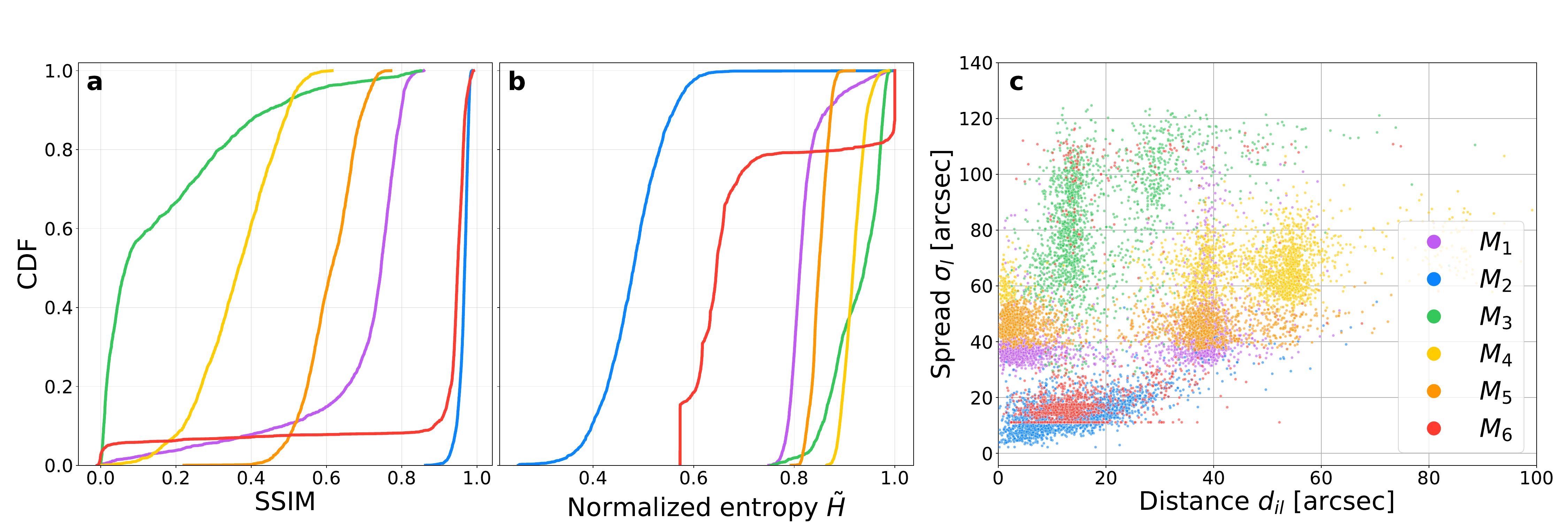}
    \caption{Grad-CAM localisation diagnostics for all six models, evaluated on $\mathbf{R}_{L1}$ test-fold sources. \textbf{(a)}~CDF of SSIM between each model's normalised heatmap and the $3\sigma$ source mask. \textbf{(b)}~CDF of normalised Shannon entropy $\tilde{H}$. \textbf{(c)}~Peak distance $d_{il}$ versus activation spread $\sigma_l$ (both in arcsec); each point represents a single source. The colour legend in panel~(c) applies to all three panels.}
    \label{fig:cam}
\end{figure*}

\subsection{Discriminative ability}

Receiver Operating Characteristic (ROC) curve plots the true positive rate against the false positive rate at varying classification thresholds, and the area under this curve (AUC-ROC) summarises a classifier's ability to separate classes across all operating points; a value of 1.0 denotes perfect separation and 0.5 corresponds to random guessing. The precision-recall (PR) curve instead plots precision against recall, and its area (AUC-PR, equivalently average precision) provides a complementary summary that is more sensitive to performance on minority classes, since it does not reward correct rejection of true negatives. Figure~\ref{fig:roc-pr} presents both curves for all six models under the two dataset regimes $\mathbf{R}_{L1}$ and $\mathbf{R}_{L2}$.

A clear performance hierarchy emerges. $M_1$ leads with an AUC-ROC of 0.95 in both regimes, followed closely by $M_2$ and $M_3$ at 0.94. $M_4$ and $M_5$ form a second tier at 0.92, while $M_6$ trails distinctly at 0.88--0.90. The same ranking is preserved in the PR curves, but the gaps widen: $M_1$ achieves an average precision of 0.88, $M_2$ and $M_3$ reach 0.85--0.87, whereas $M_4$ and $M_5$ drop to 0.81--0.83 and $M_6$ falls to 0.74--0.77. This amplification occurs because the PR metric penalises imprecise predictions on minority classes that the ROC, by crediting true negatives, can mask.

A second notable result is the near-complete overlap between the $\mathbf{R}_{L1}$ and $\mathbf{R}_{L2}$ curves for the best performing models: AUC-ROC values are identical or differ by at most 0.03, and average precisions shift by comparably small margins. This insensitivity indicates that at least five of the models have learned to disregard spurious sources, an encouraging finding for deployment on samples that have not undergone aggressive source-quality filtering. The one partial exception is $M_6$, for which $\mathbf{R}_{L1}$ slightly outperforms $\mathbf{R}_{L2}$.

\subsection{Model calibration}\label{sec:calibration}

A reliability diagram plots the fraction of positives (empirical accuracy) against the mean predicted probability (confidence) in discrete bins; a perfectly calibrated model falls on the diagonal, meaning its stated confidence matches its actual correctness rate. Departure from this diagonal is quantified by the expected calibration error (ECE), the weighted average absolute difference between confidence and accuracy across bins. Figure~\ref{fig:ece} presents reliability diagrams for all six models under both dataset regimes, averaged over 5-fold cross-validation with shaded $\pm1$ standard deviation bands; ECE values are reported in the legend.

$M_2$ is the best-calibrated model by a substantial margin, achieving the lowest ECE of $0.07\pm0.01$ ($\mathbf{R}_{L1}$) and $0.08\pm0.01$ ($\mathbf{R}_{L2}$). Its reliability curve tracks the diagonal more closely than any other architecture, particularly in the high-confidence bins where classification decisions are most consequential, and the shaded band confirms that this advantage is consistent across folds. By contrast, $M_4$ and $M_6$ are the most poorly calibrated (ECE $0.17$--$0.19$), and even strong discriminators like $M_1$ exhibit moderate miscalibration (ECE $\sim$0.11--0.12). $M_5$ achieves a low ECE ($\sim$0.11--0.12) despite weaker discriminative performance, indicating that calibration and discrimination are not always correlated.
Since standard post-hoc recalibration can close this calibration gap for the baselines, $M_2$'s native calibration is best understood as a practical convenience for workflows rather than as an irreducible advantage.

\section{Discussion} \label{sec:discussion}

\subsection{Spatial fidelity of model attention} \label{sec:attention}

To quantify how faithfully each model's Grad-CAM (\citealt{Selvaraju2017}) heatmap traces the underlying radio morphology, we define four complementary metrics. Because the test fold rotates across the five cross-validation runs, every one of the 2060 labelled sources appears in exactly one test batch, allowing us to generate Grad-CAM maps for the entire dataset. For each test-fold source, Grad-CAM is applied to the predicted class: the logit score $s_{\hat{c}}$ is back-propagated to obtain channel-wise gradient weights, which are used to produce a ReLU-activated, weighted sum of the target-layer feature maps. For $M_1$ and $M_3$--$M_5$, forward hooks capture activations at architecture-specific target layers; for $M_2$, activations are stored after the second equivariant ReLU of the DSteerableLeNet encoder. For the transformer-based architectures ($M_4$, $M_6$), patch-token activations are reshaped from sequence form to the spatial grid; the class token is excluded. In all cases the native-resolution heatmap is bilinearly resized to $150^2$ pixels and normalised to $[0,1]$.

In order to define the metrics, let $\mathbf{H}$ denote the min--max-normalised heatmap and $\mathbf{S}$ the binary source mask obtained by thresholding the grayscale radio image from $\mathbf{R}_{L1}$ or $\mathbf{R}_{L2}$ at $3\sigma$ above its mean. The first metric, the \textit{structural similarity index} (SSIM; \citealt{Wang2004}) between $\mathbf{H}$ and $\mathbf{S}$, is the mean over all local $11\times11$ windows of
\begin{equation}
\mathrm{SSIM}(x,y) = \frac{(2\mu_x\mu_y + C_1)(2\sigma_{xy} + C_2)}{(\mu_x^2 + \mu_y^2 + C_1)(\sigma_x^2 + \sigma_y^2 + C_2)},
\end{equation}
where $\mu$, $\sigma^2$, and $\sigma_{xy}$ are the local means, variances, and cross-covariance within each window, and $C_1 = (0.01\,L)^2$, $C_2 = (0.03\,L)^2$, with $L$ the dynamic range of the inputs. SSIM $\in [-1, 1]$, with values near 1 indicating that the attention map closely resembles the source morphology in luminance, contrast, and structure. The second metric, the \textit{normalised Shannon entropy} of $\mathbf{H}$, is
\begin{equation}
\tilde{H} = -\frac{1}{\ln N}\sum_{i} p_i \ln p_i,
\end{equation}
where $p_i = (H_i + \epsilon)/\sum_j(H_j + \epsilon)$ with $\epsilon = 10^{-12}$ and $N$ is the total number of pixels. $\tilde{H} \in [0,1]$, with lower values indicating more concentrated activation. To characterise where the activation falls relative to the source, we define two further metrics. The \textit{peak distance} $d_{il}$ is the Euclidean distance (in arcsec) between the heatmap peak pixel and the centre of mass of $\mathbf{S}$. The \textit{activation spread} is $\sigma_l = \sqrt{\mathrm{Var}_x + \mathrm{Var}_y}$, where $\mathrm{Var}_x = \sum_{ij} w_{ij}(x_{ij} - \bar{x})^2$ and $\mathrm{Var}_y = \sum_{ij} w_{ij}(y_{ij} - \bar{y})^2$ are the weighted spatial variances with weights $w_{ij} = H_{ij}/\sum H$, and $(\bar{x}, \bar{y})$ is the activation centroid. Both $d_{il}$ and $\sigma_l$ are converted to arcsec using the FIRST pixel scale. Small $d_{il}$ combined with moderate $\sigma_l$ indicates attention that is centred on and spread commensurate with the source extent. All four metrics are evaluated per source; the distributions of the first two are presented as empirical cumulative distribution functions (CDFs) to allow non-parametric comparison across models without binning artefacts.

Panel~(a) of Fig.~\ref{fig:cam} shows the CDF of SSIM. $M_3$ (green) rises earliest, reaching CDF $\approx 1$ by SSIM $\approx 0.5$, indicating that nearly all of its attention maps bear little structural resemblance to the actual radio source. $M_1$ (violet), $M_4$ (yellow), and $M_5$ (orange) occupy the middle range, with $M_4$ and $M_5$ showing a long tail of low-SSIM sources before rising steeply, and $M_1$ following a more gradual rise that reaches high SSIM values but with greater spread. The most striking feature is the behaviour of $M_2$ (blue) and $M_6$ (red): both curves are pushed furthest to the right, meaning these two models produce the highest-fidelity attention maps overall. From CDF $\approx 0.2$ upward, $M_6$ closely tracks $M_2$, and the two remain nearly degenerate through the upper range. Taken at face value, this would suggest that $M_2$ and $M_6$ attend to source morphology with comparable fidelity.

Panel~(b) delivers the first clue. $M_2$ has the steepest CDF rise at the lowest entropy values, reaching CDF $\approx 1$ well before any other model; nearly all of its heatmaps concentrate activation into a narrow, structured region. $M_6$, by contrast, shows a distinctive two-stage profile: a steep initial rise that plateaus around CDF $\approx 0.8$, followed by a delayed second rise toward CDF $= 1$ at higher entropy. This plateau reveals that roughly 20\% of $M_6$'s heatmaps are substantially more diffuse than the rest---a bimodality absent from $M_2$'s unimodal, consistently low-entropy distribution. The remaining models span the high-entropy end: $M_3$ and $M_4$ are the most diffuse, while $M_1$ and $M_5$ cluster together at intermediate values. Entropy alone, however, cannot distinguish structured concentration from a compact but morphologically uninformative blob, since both yield low $\tilde{H}$.

Panel~(c) resolves this ambiguity. In the $d_{il}$--$\sigma_l$ plane, $M_2$ (blue) occupies a tight, unimodal locus at low-to-moderate peak distance and low spread: its activation peaks land close to the source centre of mass and extend outward to trace the source extent. $M_6$ (red) splits into two populations. One overlaps the $M_2$ locus; the other scatters across all $d_{il}$ at high $\sigma_l$ ($\sim$80--100$''$). This bimodality maps directly onto the entropy plateau in panel~(b): the compact $M_6$ sources yield low entropy and high SSIM, while the diffuse-halo sources inflate the entropy tail and degrade structural fidelity. $M_3$ (green) fills the opposite corner---large $d_{il}$ and low-to-moderate $\sigma_l$---confirming that its high-entropy, low-SSIM behaviour reflects attention that neither centres on nor tracks the source. $M_1$, $M_4$, and $M_5$ populate the intermediate region, consistent with their mid-range SSIM and entropy distributions. The three panels together thus form a diagnostic chain: SSIM~(a) poses an apparent near-equivalence between $M_2$ and $M_6$ (visible from CDF $\approx 0.2$ upward); entropy~(b) exposes a bimodal instability in $M_6$; and panel~(c) identifies its origin---$M_6$ oscillates between a compact on-source blob and a diffuse halo, whereas $M_2$ consistently traces source morphology.

The figures of Appendix \ref{app:cam} illustrate Grad-CAM contours overlaid on the original images ($\mathbf{R}_{L1}$) for eight representative sources for each morphological class. A clear architectural divide emerges. The CNN-based supervised models $M_1$, $M_3$, and $M_5$ behave similarly: their contours partially overlap the source but default to smooth, elliptical patterns that lack fine morphological detail, consistent with their low SSIM and high entropy distributions in panels~(a) and~(b). Among the non-CNN architectures, $M_4$ (Swin) spreads attention across the entire image with broad, loosely spaced contours, evident from the spread of yellow dots in Panel~(c), while $M_6$ (ViT) over-concentrates into compact concentric rings centred on the brightest pixel, capturing the core but discarding extended structure---the two $M_6$ populations identified in panel~(c) are directly visible here as sources where the contours either tightly ring the peak or diffuse across the field. $M_2$ is the outlier: its contours closely trace lobes, jets, and bends with tightly nested isocontours that follow the source boundary rather than merely enclosing it, explaining its consistently low entropy and highest SSIM. The contrast is most pronounced for bent and extended sources (e.g.\ L0809, L2121), where $M_2$'s contours bend with the source while all other models default to circular or image-filling patterns.

This morphology-tracing behaviour of $M_2$ is established early in training and is not the product of extended fine-tuning: as shown in Appendix~\ref{app:epochs} (Figs.~\ref{fig:ssim-H} and~\ref{fig:loc-epochs}), $M_2$'s attention is stable from the earliest checkpoint, while the baselines remain pinned at their respective plateaus and $M_6$ destabilises. The more plausible drivers are therefore architectural---BYOL pre-training on unlabelled radio sources and the equivariant E2CNN encoder.

\begin{table}
\centering
\small
\caption{Class distributions for the five imbalance experiments. All experiments use 1288 sources. The Gini coefficient measures class imbalance, with 0 indicating perfect balance.}
\label{tab:imbalance_experiments}
\begin{tabular}{llrrrrrc}
\hline\hline
ID & Imbalance & sFRI & sFRII & WAT & NAT & Total & Gini \\
\hline
$E_1$ & Balanced         & 322 & 322 & 322 & 322 & 1288 & 0.000 \\
$E_2$ & Mild             & 303 & 323 & 357 & 305 & 1288 & 0.035 \\
$E_3$ & Moderate         & 257 & 339 & 371 & 321 & 1288 & 0.070 \\
$E_4$ & High             & 246 & 368 & 396 & 278 & 1288 & 0.105 \\
$E_5$ & Severe           & 208 & 385 & 414 & 281 & 1288 & 0.140 \\
\hline
\end{tabular}
\end{table}

\begin{figure}
    \centering
    \includegraphics[width=\columnwidth]{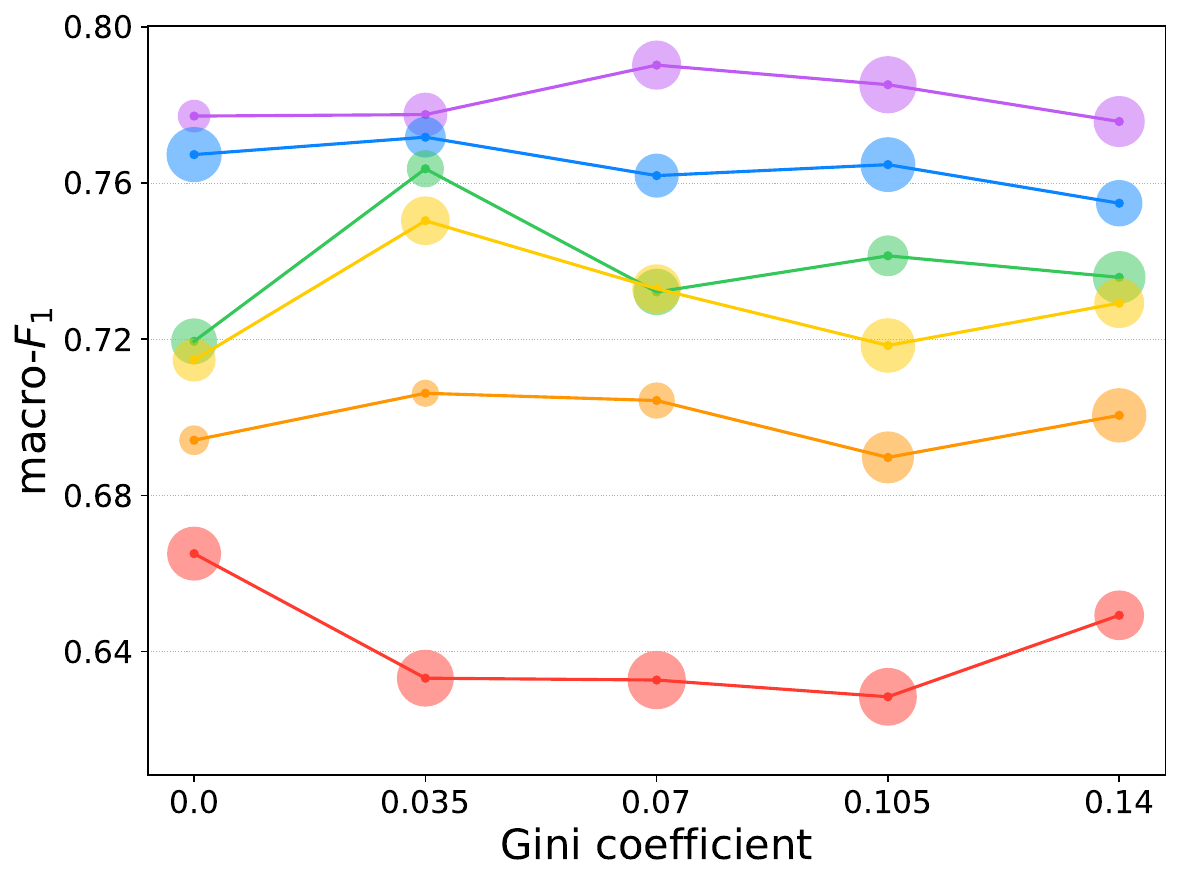}
    \caption{Macro-$F_1$ versus Gini coefficient for all six models under $\mathbf{R}_{L1}$, corresponding to experiments $E_1$--$E_5$ (Table~\ref{tab:imbalance_experiments}). Bubble diameter is proportional to the standard deviation across 5-fold cross-validation. Colours follow Fig.~\ref{fig:cam}.}
    \label{fig:gini}
\end{figure}

\subsection{Sensitivity to class imbalance} \label{sec:imbalance}

To assess robustness to class imbalance, we construct five datasets ($E_1$--$E_5$) from $\mathbf{R}_{L1}$, each containing 1288 sources but with progressively skewed class distributions characterised by evenly spaced Gini coefficients from 0.000 (perfectly balanced) to 0.140 (severe imbalance), as listed in Table~\ref{tab:imbalance_experiments}. The class ordering WAT $\geq$ sFRII $\geq$ NAT $\geq$ sFRI is maintained across all experiments, with a floor of 200 sources per class to avoid degenerate training. By fixing the total sample size, any change in performance can be attributed to the distributional shift rather than to a reduction in training data. All six models are trained and evaluated on each dataset using 5-fold cross-validation.

Fig.~\ref{fig:gini} shows macro-$F_1$ as a function of Gini coefficient for all six models, with bubble size indicating the standard deviation across folds. A mild degree of imbalance (Gini $\approx 0.035$) improves performance for five of the six models relative to the perfectly balanced case; the exception is $M_6$, which drops. $M_1$ and $M_2$ are the most robust, remaining consistently above 0.75 across the entire Gini range, with $M_1$ peaking at Gini $\approx 0.07$. $M_3$ and $M_4$ track each other closely at an intermediate level ($\sim$0.72--0.75), though $M_4$ dips more noticeably at high Gini while $M_3$ holds steady. $M_5$ shows stability comparable to $M_1$ and $M_2$ but at a lower plateau ($\sim$0.70). $M_6$ is consistently the weakest, fluctuating around 0.64. The ranking thus partially reshuffles under imbalance: at perfect balance the ordering is $M_1 > M_2 > M_3 \approx M_4 > M_5 > M_6$, but at severe imbalance $M_4$ drops toward $M_5$ while $M_1$, $M_2$, $M_3$, and $M_5$ hold their positions.

Fig.~\ref{fig:imbalance} separates the results into correct classification rate, within-type error (confusions between morphologically similar classes: sFRI$\leftrightarrow$sFRII or WAT$\leftrightarrow$NAT), and cross-type error (confusions between straight and bent sources). The correct-classification panel preserves the familiar model ranking across all five experiments, with $M_1$ and $M_2$ showing very small variation with imbalance level. The error decomposition reveals that these two top models achieve their similar overall scores through different strengths: $M_2$ achieves the lowest within-type error across nearly every experiment, confirming that its morphology-tracing attention translates into superior discrimination between sources differing only in fine structural detail, while $M_1$ achieves the lowest cross-type error, suggesting that its ImageNet-pretrained features are particularly effective at capturing the gross straight-versus-bent distinction.
This effect has been demonstrated for RAGNs by \citet{Tang2019} and is consistent with the layered structure of CNN features: early ImageNet layers learn generic edge, curve, and orientation detectors \citep{Cammarata2020} that respond to elongated and bent shapes regardless of semantic identity, while only the highest semantic features are re-learned during fine-tuning \citep{Yosinski2014}. The complementary strengths thus suggest that $M_1$ and $M_2$ have learned partially independent feature representations.

\subsection{Astrophysical implications} \label{sec:astro}

Bent RAGNs have long served as tracers of dense environments. \citet{Garon2019} showed that 87\% of a sample of over 4000 RAGNs from the RGZ lie within 50~Mly of an optically identified cluster, and \citet{Golden-Marx2021,GoldenMarx2023} demonstrated that sources with narrower opening angles reside in richer clusters. Our four-class scheme adds a dimension that existing binary or three-class classifiers lack: by resolving bent sources into WATs and NATs rather than treating them as a single undifferentiated class, it enables a first-order separation between the BCGs that typically host WATs \citep{ODea2023, Burns1998} and the satellite galaxies that typically host NATs \citep{miley1972, rudnick1976}. WATs are the most abundant bent class in our dataset, consistent with their association with the dominant, centrally located galaxies in groups and clusters. Interestingly, NATs achieve the highest per-class $F_1$-scores across all models in Fig.~\ref{fig:f1-heatmap}, despite being less numerous. This may reflect their more morphologically distinctive appearance: the closely collimated, often single-tailed structure of NATs presents a sharper contrast with the straight FR classes than the broader, more symmetric WAT morphology does, which can more easily be confused with sFRI sources. Applied at scale to EMU, LoTSS, and ultimately the SKA, a WAT/NAT-resolved catalogue would provide not just cluster candidates but a morphological proxy for the host galaxy's dynamical state within the cluster.

The morphology of radio tails also carries quantitative information about the ICM itself. Recent work has demonstrated strong correlations between jet bending angle, cluster-centric distance, and galaxy velocity \citep{vanderJagt2025}, while spectral aging along bent tails constrains both the radiative losses of the relativistic electron population and the infall history of the host galaxy. The Grad-CAM analysis of Section~\ref{sec:attention} is relevant here: $M_2$ is the only model whose attention contours consistently trace bends, lobes, and extended tails (Appendix~\ref{app:cam}) rather than defaulting to compact central blobs ($M_6$) or diffuse image-filling patterns ($M_4$). A classifier that attends to the morphological features carrying ICM information---bending curvature, jet--tail transitions, lobe asymmetry---is a necessary condition for any future attempt to extract quantitative environmental information from classifier internals at survey scale.

Finally, tailed radio galaxies are increasingly recognised as progenitor sources of diffuse cluster radio emission \citep{vanWeeren2019, Vazza2024}. The fossil plasma deposited by AGN tails into the ICM can be re-energized by merger shocks or turbulence to produce radio phoenices and gently re-energized tails \citep{vanWeeren2019, deGasperin2017, Mandal2020}. The morphological class of the progenitor is expected to influence where and how this fossil plasma is distributed: straight sources deposit energy along the jet axis, potentially at large distances from the cluster centre, while bent tails are more likely to distribute relativistic plasma within the cluster volume itself \citep{ODea2023, Vazza2024}. Our sFRI/sFRII versus WAT/NAT classification is therefore relevant to the fossil plasma supply chain---automated classification at survey scale would enable statistical studies of which progenitor types seed which classes of diffuse emission.

\section{Conclusions} \label{sec:conclusions}

We present FIRST-2060, a four-class labelled dataset of 2060 radio AGNs classified as sFRI, sFRII, WAT, and NAT (Section~\ref{sec:data}, Table~\ref{tab:data}), and RGC 1.0, a semi-supervised model combining BYOL pre-training with an E2CNN encoder, fine-tuned for the four-class problem (Section~\ref{sec:models}, Table~\ref{tab:models}). We benchmark RGC 1.0 ($M_2$) against five supervised baselines ($M_1$, $M_3$--$M_6$) initialised from ImageNet.

Before summarising the principal findings, we note that the macro-F1 of our four-class model is substantially lower than the ${>}\,97\%$ accuracy of the binary FRI/FRII classifier in H23. This gap is explained by the four-class landscape itself: the confusion matrices (Fig.~\ref{fig:conf1}) show that errors are dominated by within-type confusions (sFRI$\leftrightarrow$sFRII, WAT$\leftrightarrow$NAT) that did not exist in the binary setting, and the imbalance experiments (Fig.~\ref{fig:imbalance}) confirm that even perfect class balance yields no model above macro-F1 ${\sim}\,0.78$, placing the ceiling at morphological ambiguity rather than distributional skew. The principal findings of this work are as follows.

\begin{enumerate}

\item The model ranking $M_1 > M_2 > M_3 > M_4 > M_5 > M_6$ in macro-$F_1$ and accuracy is consistent across both dataset regimes, $\mathbf{R}_{L1}$ and $\mathbf{R}_{L2}$ (Fig.~\ref{fig:f1-heatmap}). $M_1$ and $M_2$ form a top tier whose internal difference ($0.80 \pm 0.02$ vs.\ $0.79 \pm 0.02$) is within one standard deviation, separated from the remaining models by more than one standard deviation.

\item $M_2$ achieves this near-parity despite having fewer parameters (53M vs.\ 89M) and being pretrained on 20\,000 unlabelled radio sources rather than ImageNet. It also incorporates rotational equivariance as a physical prior, encoding the orientation-independence of morphological class.

\item $M_2$ is the best-calibrated model (ECE $\approx 0.07$; Fig.~\ref{fig:ece}), substantially ahead of the next-best architecture. Well-calibrated probability estimates matter for downstream tasks such as confidence-thresholded catalogue construction and Bayesian source characterisation.

\item NATs are the most reliably classified class across all models, while sFRI sources are consistently the hardest (Fig.~\ref{fig:f1-heatmap}). The confusion matrices (Fig.~\ref{fig:conf1}) confirm that the error budget is dominated by within-type confusion---sFRI$\leftrightarrow$sFRII and WAT$\leftrightarrow$NAT---rather than cross-type errors between the straight and bent families.

\item Five of the six models show near-identical discriminative performance between $\mathbf{R}_{L1}$ and $\mathbf{R}_{L2}$ (Fig.~\ref{fig:roc-pr}), indicating that they have learned to disregard spurious sources. The exception is $M_6$, whose WAT performance degrades on $\mathbf{R}_{L2}$.

\item Grad-CAM analysis (Section~\ref{sec:attention}, Fig.~\ref{fig:cam}) shows that $M_2$ is the only model whose attention contours consistently trace the morphological structure of RAGNs---lobes, jets, and bends---rather than defaulting to compact blobs or diffuse patterns (Appendix~\ref{app:cam}). Because these structures encode the physical interaction between the AGN and the surrounding medium, spatially faithful attention is a necessary condition for any future attempt to extract quantitative environmental information from classifier internals.

\item $M_1$ and $M_2$ are the most robust to class imbalance (Fig.~\ref{fig:gini}), but error decomposition (Fig.~\ref{fig:imbalance}) reveals complementary strengths: $M_1$ excels at cross-type separation (straight vs.\ bent), while $M_2$ excels at within-type discrimination (sFRI vs.\ sFRII, WAT vs.\ NAT). An ensemble or hierarchical classifier combining both could exploit these complementary profiles.

\end{enumerate}

The four-class scheme introduced here---resolving bent sources into WAT and NAT rather than a single undifferentiated class---is directly applicable to the ongoing EMU and LoTSS surveys and to the forthcoming SKA. Automated WAT/NAT-resolved catalogues would provide not only cluster candidates but also a morphological proxy for the dynamical state of the host galaxy within its environment. Scaling the pre-training to larger unlabelled radio datasets, following \citet{Slijepcevic2024}, is a natural next step toward building domain-specific foundation models for radio astronomy.

\section*{Data and Code Availability}
The FIRST-2060 dataset, trained model weights, and all code used in this study are publicly available in our GitHub repository: \url{https://github.com/cassaiub/rgc}.

\section*{Acknowledgments}
Asad, Amin, and Momen acknowledge support from ICT Innovation Fund Grant No.~148 (2020) of the ICT Division, Government of Bangladesh, and from Sponsored Research Grant No.~2020-SETS-13 of Independent University, Bangladesh (IUB). We are especially grateful to the anonymous referee, whose detailed and incisive report motivated us to transform this work from a binary WAT/NAT classifier into the substantially broader four-class benchmark presented here; the earlier, narrower version is available as \texttt{arXiv:2510.22190v1}. The authors used Claude and Gemini as writing and coding aids to refine portions of the manuscript for clarity and readability. All scientific content, analysis, and conclusions are entirely the authors' own; the authors reviewed and edited all AI-assisted text and take full responsibility for the final version of the manuscript.

\section*{Contributor Roles Taxonomy (CRediT)}
The first five authors are listed in order of significance of their contribution; the last author is the supervisor of the first author. All other authors are listed alphabetically. Author contributions are described using the taxonomy of \citealt{Brand2015}.
\textit{Hossain}: Conceptualization;
\textit{Shahal}: Software, Validation, Data Curation, Visualization;
\textit{Asad}: Writing -- Original Draft, Writing -- Review \& Editing, Supervision;
\textit{Saikia}: Conceptualization, Writing -- Review \& Editing;
\textit{Khan}: Data Curation;
\textit{Akter}: Data Curation;
\textit{Ali}: Supervision;
\textit{Amin}: Funding Acquisition;
\textit{Guha}: Data Curation;
\textit{Jihad}: Data Curation;
\textit{Momen}: Resources;
\textit{Sen}: Data Curation;
\textit{Rahman}: Supervision, Project Administration.

\bibliographystyle{aa}
\bibliography{refs}

%-------------------------------------------------------------
% Appendix
%-------------------------------------------------------------
\begin{appendix}

\onecolumn
\section{Confusion matrices} \label{app:conf}

\begin{figure}[htbp]
    \centering
    \includegraphics[width=\linewidth]{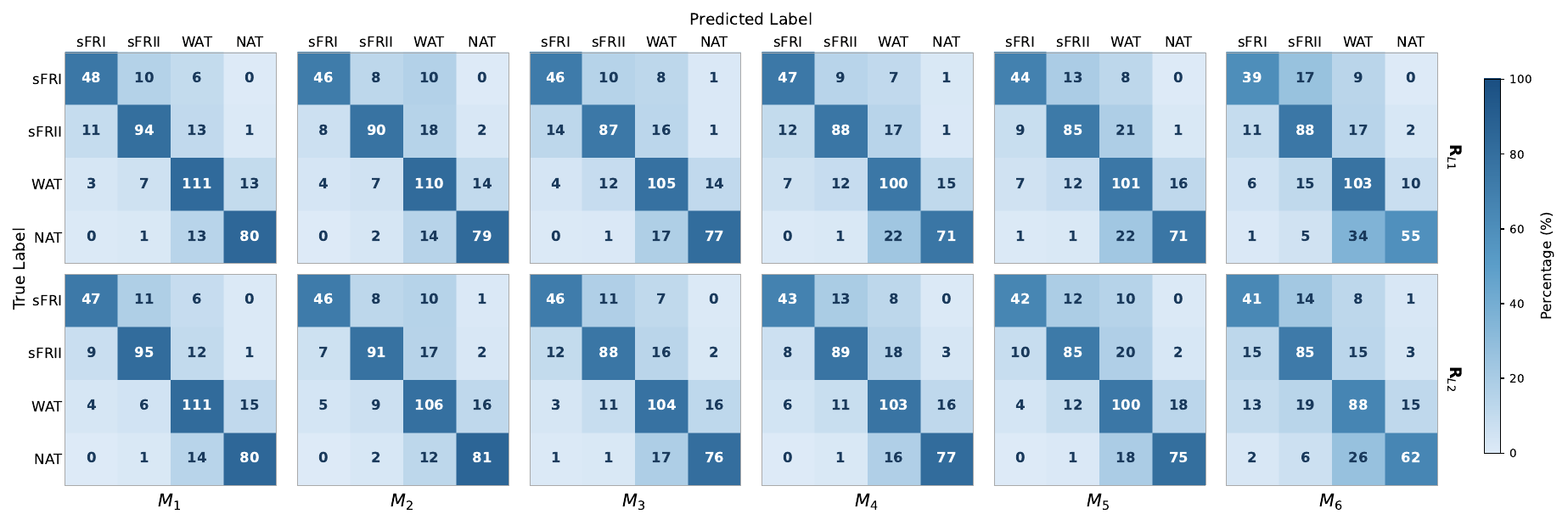}
    \caption{Confusion matrices for the six evaluated models ($M_1$–$M_6$) across the four morphological classes. Results are shown for both dataset regimes: $\mathbf{R}_{L1}$ (spurious sources retained) and $\mathbf{R}_{L2}$ (spurious sources removed). Cell values denote the mean absolute number of sources per test fold, averaged over the 5-fold cross-validation. To account for class imbalance, the colour scale maps the row-wise percentage (Recall), from 0\% (light blue) to 100\% (dark blue). Across most architectures, the darkest off-diagonal elements visually confirm that within-type confusions---between sFRI$\leftrightarrow$sFRII and WAT$\leftrightarrow$NAT---constitute the primary source of classification error.}
\label{fig:conf1}
\end{figure}

\begin{figure}[htbp]
    \centering
    \includegraphics[width=0.95\linewidth]{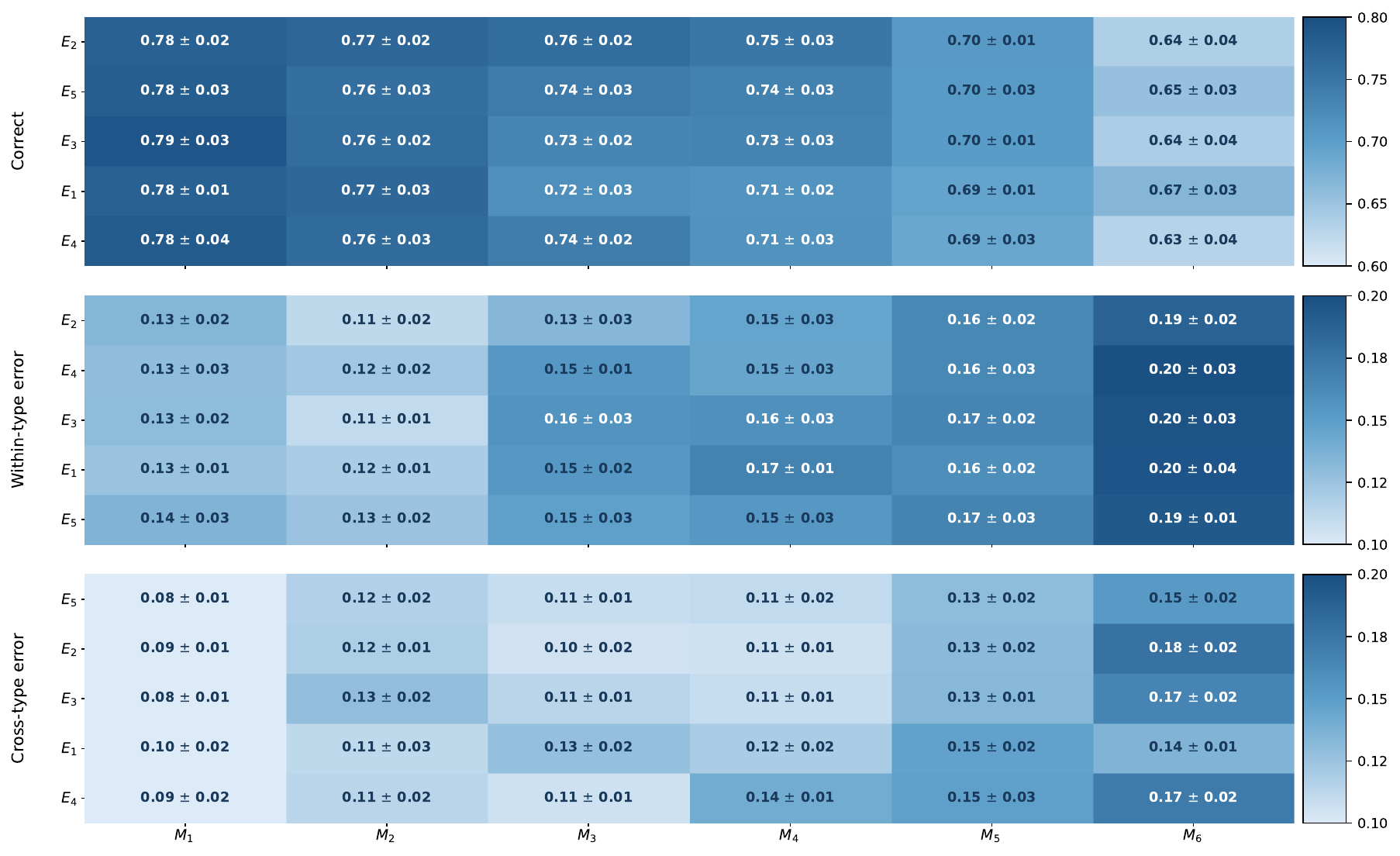}
\caption{Decomposition of classification errors across the five imbalance experiments ($E_1$--$E_5$, rows) and six models ($M_1$--$M_6$, columns). The top panel shows the correct classification rate, the middle panel the within-type error rate (confusion within the straight family sFRI/sFRII or within the bent family WAT/NAT), and the bottom panel the cross-type error rate (confusion between the straight and bent families). In each panel, darker shading indicates higher values on the respective colourbar. Values are macro-averaged over the four classes and reported as mean $\pm$ standard deviation over 5-fold cross-validation.}
    \label{fig:imbalance}
\end{figure}

\clearpage

\section{FIRST-2060 catalogue}

\small
\renewcommand{\arraystretch}{1.1}

\begin{longtable}{lllll}
\caption{Sample of the FIRST-2060 catalogue constructed for this study. The prefix letter in the \texttt{ID} column indicates the source catalogue of the image: `S' for Sasmal, `L' for Lao, and `M' for MiraBest. The remaining columns provide the standard FIRST radio designation, the right ascension (RA) and declination (Dec) in J2000 coordinates, and the true morphological label assigned to the source. The full catalogue of 2060 sources is available in a machine-readable format.} \\
\hline
ID & Name & RA & Dec & True label \\
\hline
\endfirsthead

\hline
ID & Name & RA & Dec & True label \\
\hline
\endhead

\hline
\multicolumn{5}{r}{Continued on next page} \\
\endfoot

\hline
\endlastfoot

S0479 & \texttt{J113131.07+012827.8} & \texttt{11~31~31.07} & \texttt{+01~28~27.8} & \textbf{nat} \\
L0809 & \texttt{J082443.02+520027.1} & \texttt{08~24~43.02} & \texttt{+52~00~27.1} & \textbf{nat} \\
L0125 & \texttt{J005946.74+102240.8} & \texttt{00~59~46.74} & \texttt{+10~22~40.8} & \textbf{nat} \\
S0533 & \texttt{J134936.07+482844.0} & \texttt{13~49~36.07} & \texttt{+48~28~44.0} & \textbf{nat} \\
L2121 & \texttt{J112658.33+635639.4} & \texttt{11~26~58.33} & \texttt{+63~56~39.4} & \textbf{nat} \\
L1908 & \texttt{J105253.02+532738.4} & \texttt{10~52~53.02} & \texttt{+53~27~38.4} & \textbf{nat} \\
L1707 & \texttt{J102620.82+303556.0} & \texttt{10~26~20.82} & \texttt{+30~35~56.0} & \textbf{nat} \\
L1065 & \texttt{J085758.80+170334.3} & \texttt{08~57~58.80} & \texttt{+17~03~34.3} & \textbf{nat} \\

$\cdots$ & $\cdots$ & $\cdots$ & $\cdots$ & $\cdots$ \\

L0906 & \texttt{J083756.37+364913.3} & \texttt{08~37~56.37} & \texttt{+36~49~13.3} & \textbf{wat} \\
S0197 & \texttt{J121345.43-032709.5} & \texttt{12~13~45.43} & \texttt{-03~27~09.5} & \textbf{wat} \\
L0802 & \texttt{J082258.30+143547.5} & \texttt{08~22~58.30} & \texttt{+14~35~47.5} & \textbf{wat} \\
S0269 & \texttt{J141752.89+011442.7} & \texttt{14~17~52.89} & \texttt{+01~14~42.7} & \textbf{wat} \\
L4816 & \texttt{J232350.46+034832.8} & \texttt{23~23~50.46} & \texttt{+03~48~32.8} & \textbf{wat} \\
L0332 & \texttt{J024243.01-031144.3} & \texttt{02~42~43.01} & \texttt{-03~11~44.3} & \textbf{wat} \\
L0304 & \texttt{J022325.70+064032.8} & \texttt{02~23~25.70} & \texttt{+06~40~32.8} & \textbf{wat} \\
L0867 & \texttt{J083228.01+003907.8} & \texttt{08~32~28.01} & \texttt{+00~39~07.8} & \textbf{wat} \\

$\cdots$ & $\cdots$ & $\cdots$ & $\cdots$ & $\cdots$ \\

M0159 & \texttt{J091225.08+534139.2} & \texttt{09~12~25.08} & \texttt{+53~41~39.2} & \textbf{sfri} \\
M0214 & \texttt{J155721.40+544016.0} & \texttt{15~57~21.40} & \texttt{+54~40~16.0} & \textbf{sfri} \\
M0986 & \texttt{J110140.09+331720.6} & \texttt{11~01~40.09} & \texttt{+33~17~20.6} & \textbf{sfri} \\
M1266 & \texttt{J132000.20+253243.9} & \texttt{13~20~00.20} & \texttt{+25~32~43.9} & \textbf{sfri} \\
L1069 & \texttt{J085814.32+505333.3} & \texttt{08~58~14.32} & \texttt{+50~53~33.3} & \textbf{sfri} \\
M1162 & \texttt{J112924.14+210113.9} & \texttt{11~29~24.14} & \texttt{+21~01~13.9} & \textbf{sfri} \\
M0034 & \texttt{J144724.04-002411.0} & \texttt{14~47~24.04} & \texttt{-00~24~11.0} & \textbf{sfri} \\
M1096 & \texttt{J090234.90+204417.9} & \texttt{09~02~34.90} & \texttt{+20~44~17.9} & \textbf{sfri} \\

$\cdots$ & $\cdots$ & $\cdots$ & $\cdots$ & $\cdots$ \\

M0836 & \texttt{J134503.03+094724.0} & \texttt{13~45~03.03} & \texttt{+09~47~24.0} & \textbf{sfrii} \\
M0182 & \texttt{J104548.50+042032.6} & \texttt{10~45~48.50} & \texttt{+04~20~32.6} & \textbf{sfrii} \\
M0933 & \texttt{J113857.71+372837.0} & \texttt{11~38~57.71} & \texttt{+37~28~37.0} & \textbf{sfrii} \\
M0359 & \texttt{J134423.48-022239.3} & \texttt{13~44~23.48} & \texttt{-02~22~39.3} & \textbf{sfrii} \\
L0105 & \texttt{J005238.46+013727.0} & \texttt{00~52~38.46} & \texttt{+01~37~27.0} & \textbf{sfrii} \\
L4840 & \texttt{J233623.45+000233.6} & \texttt{23~36~23.45} & \texttt{+00~02~33.6} & \textbf{sfrii} \\
M0843 & \texttt{J141947.90+081423.4} & \texttt{14~19~47.90} & \texttt{+08~14~23.4} & \textbf{sfrii} \\
M0543 & \texttt{J094702.73+094340.2} & \texttt{09~47~02.73} & \texttt{+09~43~40.2} & \textbf{sfrii} \\

\end{longtable}

\normalsize

%-------------------------------------------------------------

\vspace{1em}

\section{Localisation maps} \label{app:cam}

The following four pages present Grad-CAM localisation maps for representative sources from the four morphological classes. To accommodate page constraints, we restrict this visualisation to the $\mathbf{R}_{L1}$ dataset regime. The maps are organised by class in the following sequence: sFRI, sFRII, WAT and NAT (one class per page).

Each page displays eight sources arranged in rows, with the corresponding catalogue index indicated along the left margin. For a given source, the six columns display the attention maps generated by the evaluated models ($M_1$--$M_6$). 

Model attention is visualised as iso-contours overlaid on the input images. Tightly nested, higher-level contours correspond to stronger network activations, highlighting the most salient regions driving the model's prediction. Overlaying these contours directly onto the input images provides a clear visual diagnostic of how faithfully each model's focus aligns with the actual physical structure of the radio source.

\begin{figure}[htbp]
    \centering
    \includegraphics[width=0.97\linewidth]{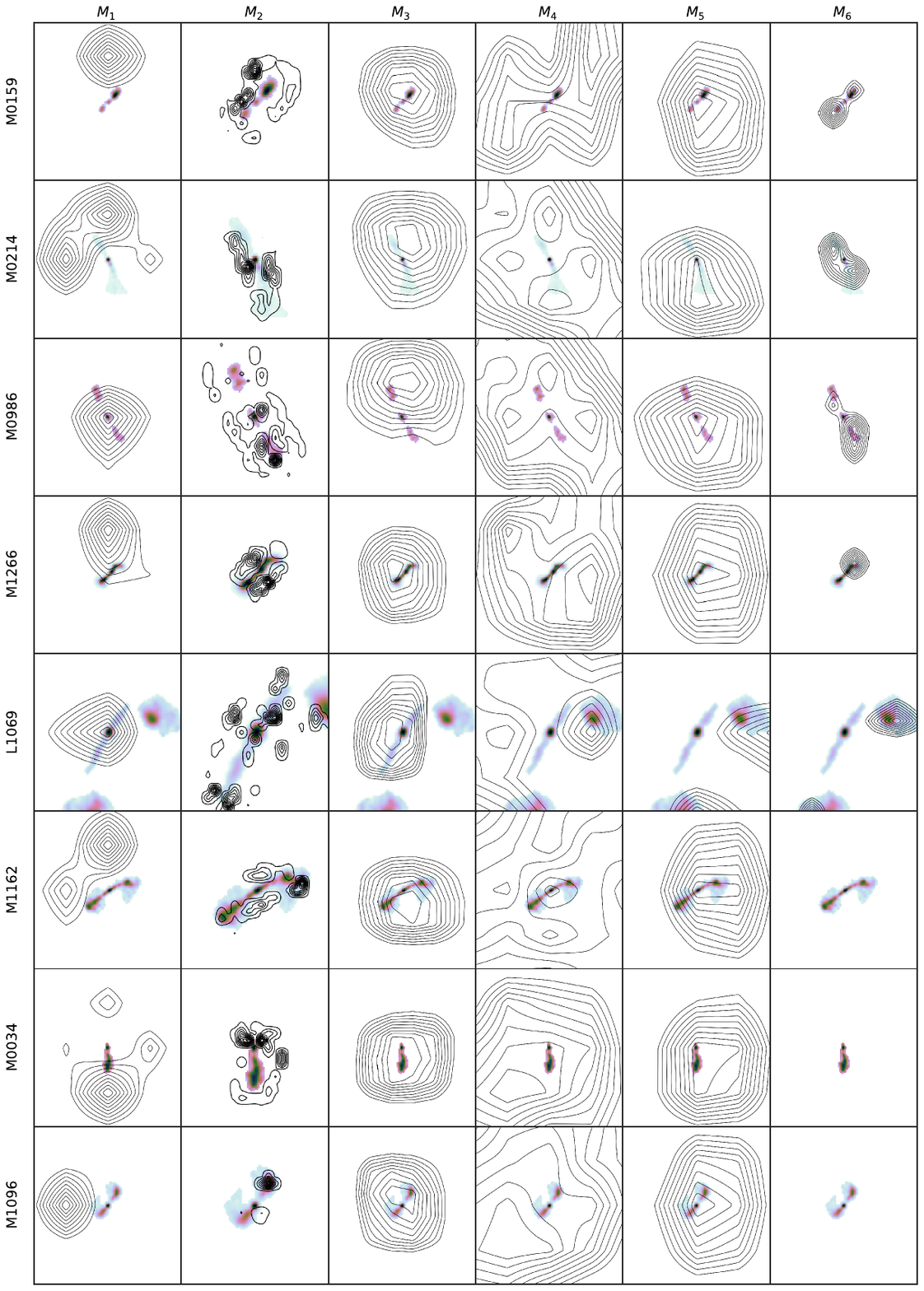}

    \label{fig:loc_sfri}
\end{figure}

\begin{figure}[htbp]
    \centering
    \includegraphics[width=0.97\linewidth]{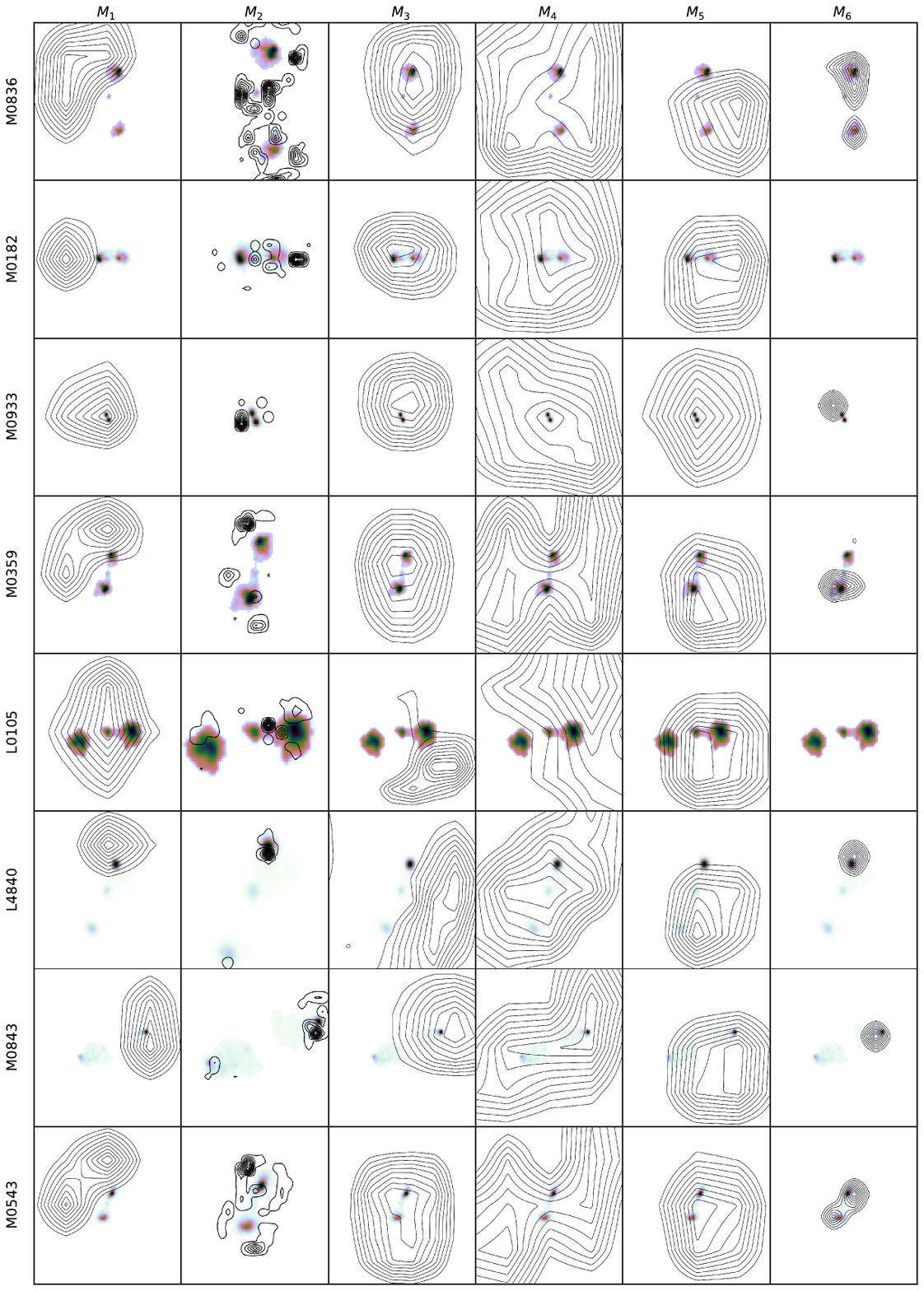}

    \label{fig:loc_sfrii}
\end{figure}

\begin{figure}[htbp]
    \centering
    \includegraphics[width=0.97\linewidth]{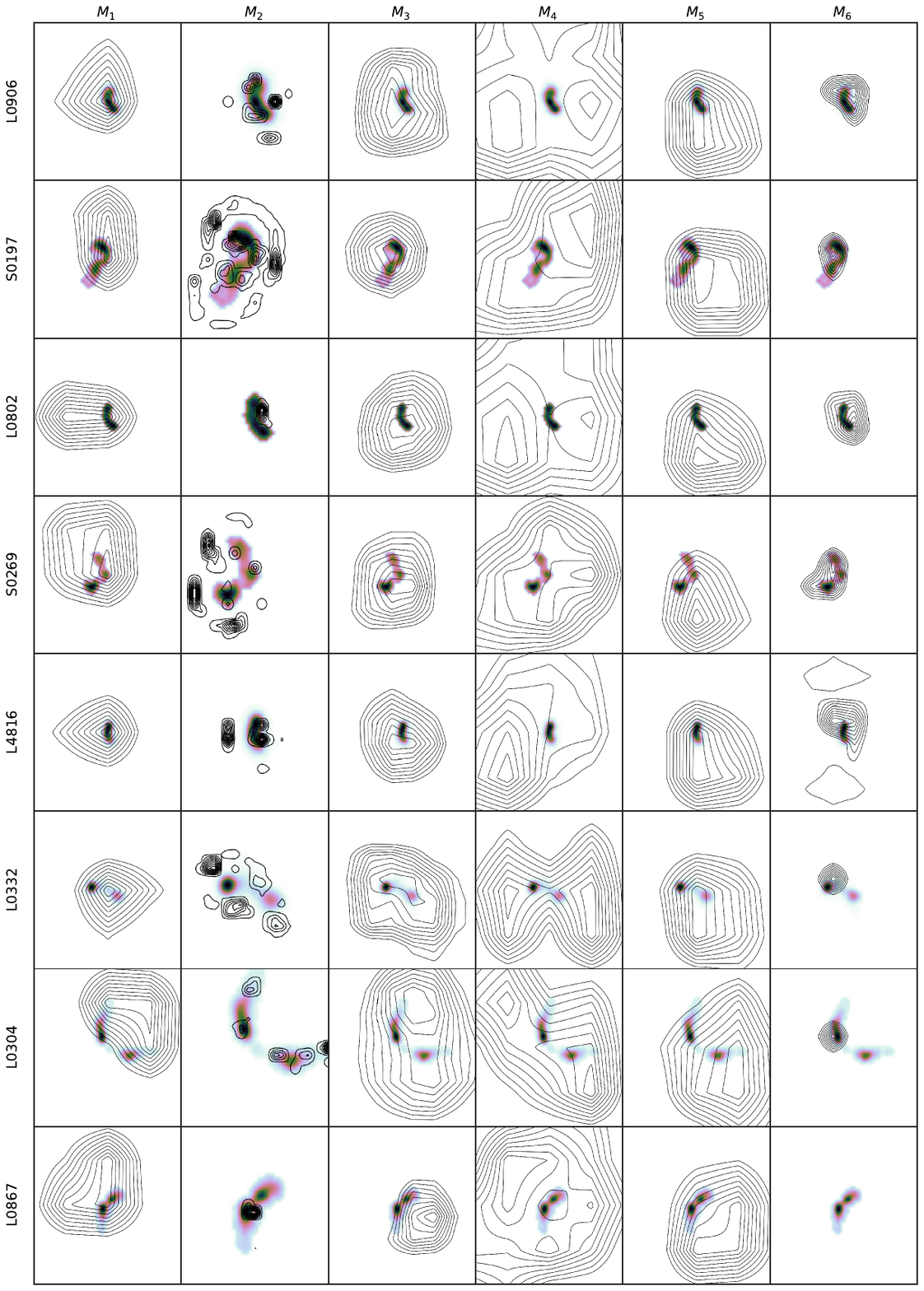}

    \label{fig:loc_wat_clean}
\end{figure}

\begin{figure}[htbp]
    \centering
    \includegraphics[width=0.97\linewidth]{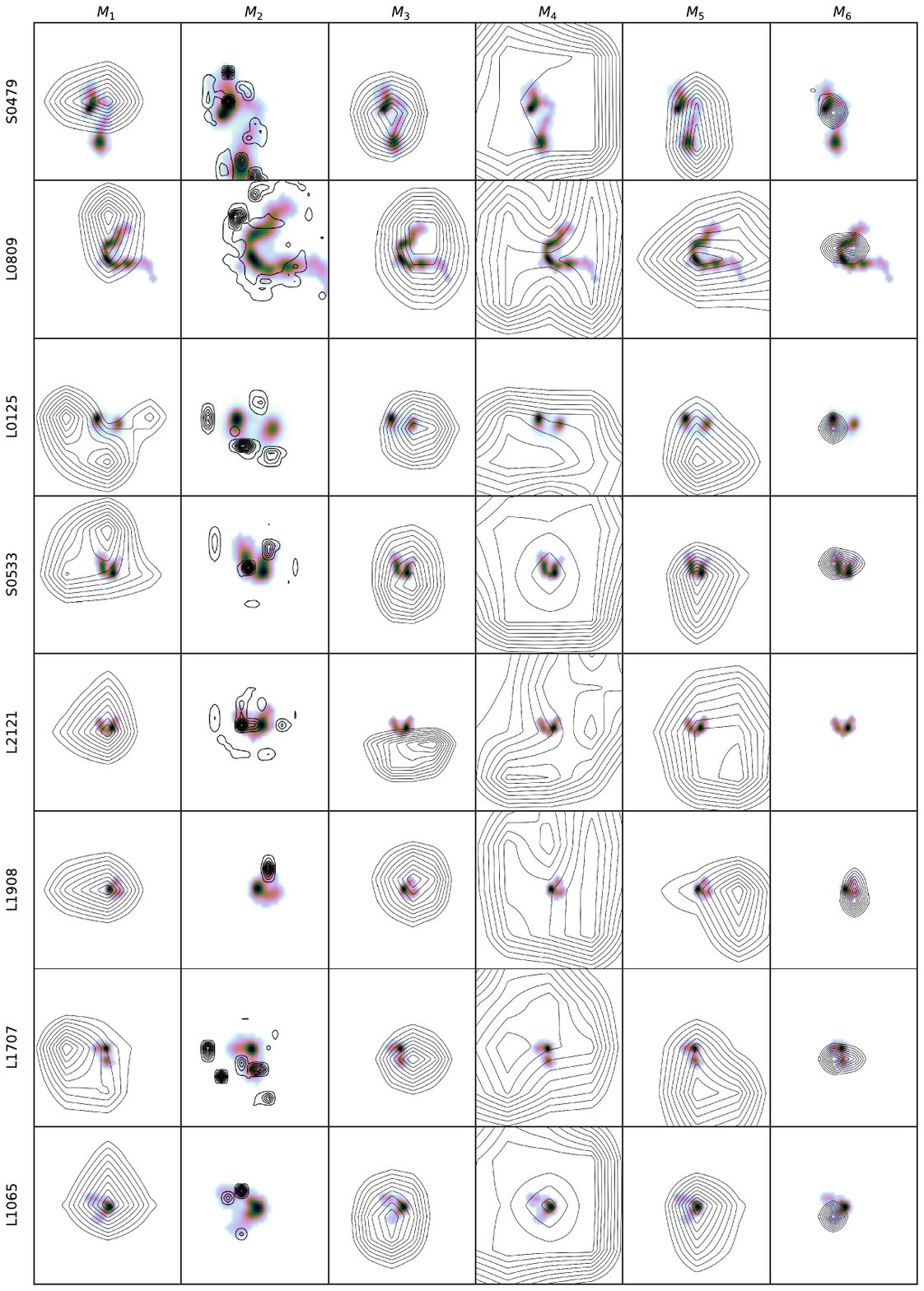}

    \label{fig:loc_nat}
\end{figure}

\clearpage

\section{Evolution across training epochs} \label{app:epochs}

\begin{figure}[htbp]
    \centering
    \includegraphics[width=0.6\linewidth]{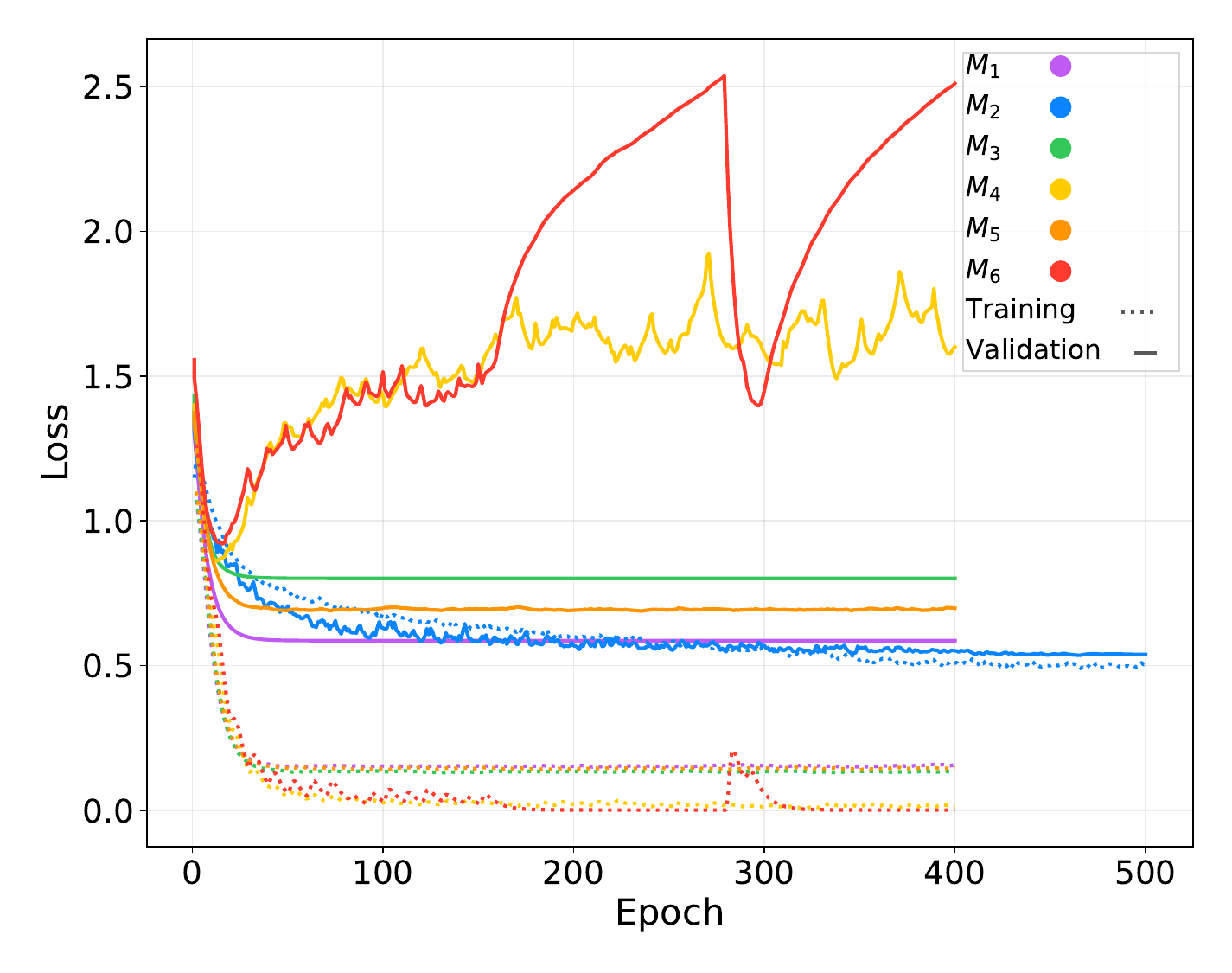}
    \caption{Training (dotted) and validation (solid) loss curves for all six models, shown over 400 epochs for the
 baselines and 500 epochs for $M_2$; all curves are smoothed using an exponential moving average (EMA) for visual
 clarity. $M_1$ converges smoothly with a small train--validation gap, and $M_2$ is the only model whose training and
  validation losses converge together (to $\sim 0.55$), indicating no overfitting. $M_3$ and $M_5$ exhibit moderate
 overfitting, with validation plateaus of $\sim 0.7$--$0.8$ while training loss approaches zero. $M_4$ and $M_6$
 overfit severely: their validation losses rise above $1.5$, and $M_6$ peaks near $2.5$ around epoch 280 with a brief
  recovery before climbing again.}

    \label{fig:loss}
\end{figure}

\begin{figure}[htbp]
    \centering
     \includegraphics[width=1\linewidth]{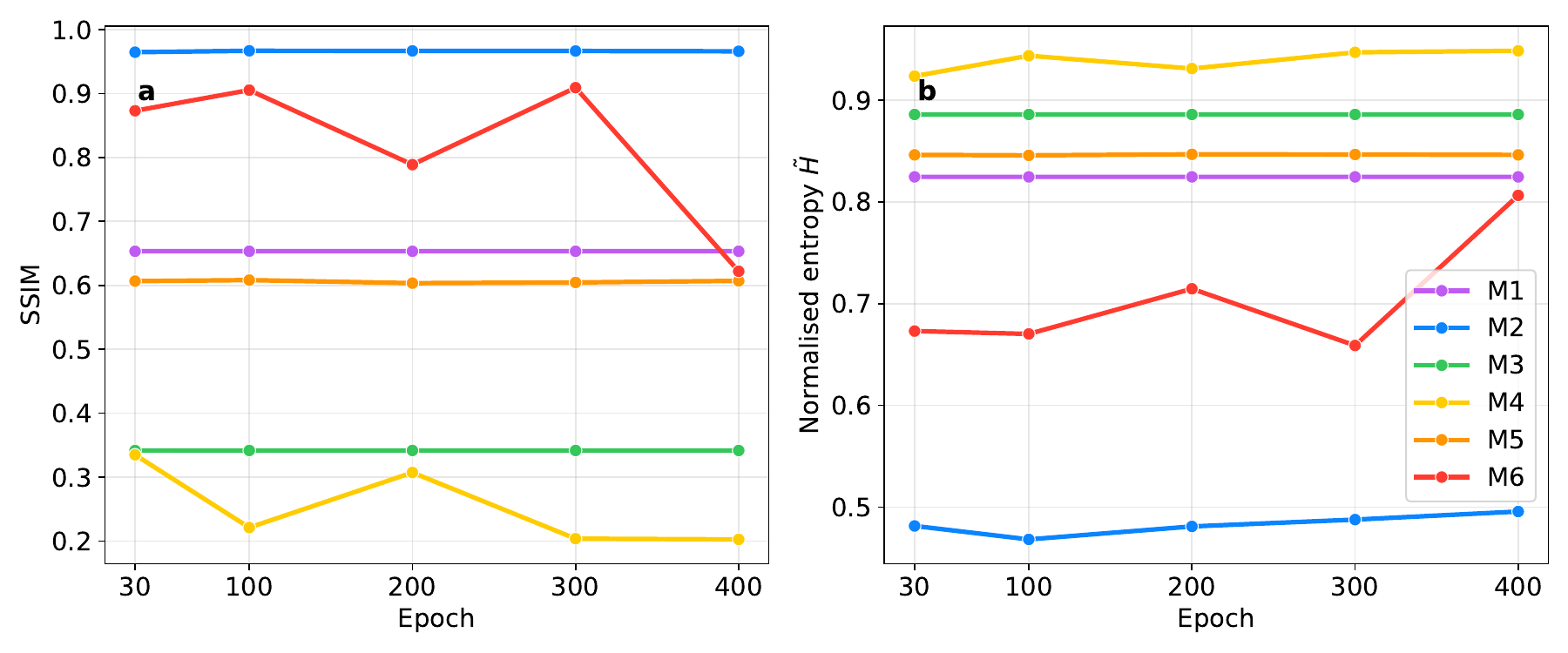}
     \caption{Evolution of Grad-CAM attention quality across training epochs for all six models ($M_1$--$M_6$; see Table~\ref{tab:models}), evaluated at checkpoints saved at epochs 30, 100, 200, 300, and 400. (a) Mean SSIM between the normalised attention heatmap and the source mask; higher values indicate greater spatial overlap. (b) Mean normalised Shannon entropy $\tilde{H}$ of the heatmap; lower values indicate more focused, spatially concentrated attention. $M_2$ dominates both panels with the highest SSIM ($\sim 0.97$) and lowest entropy ($\sim 0.48$), essentially unchanged across training. $M_1$, $M_3$, and $M_5$ remain flat at intermediate or low values, indicating that further training neither sharpens nor degrades their attention. $M_4$ (Swin-B) oscillates around the lowest SSIM throughout, while $M_6$ (ViT-B/16) is the most volatile, with SSIM falling from $\sim 0.87$ at epoch 30 to $\sim 0.62$ at epoch 400 and entropy rising in parallel.}
    \label{fig:ssim-H}
\end{figure}
\newpage

\begin{figure}[htbp]
    \centering
    \includegraphics[width=0.9\linewidth]{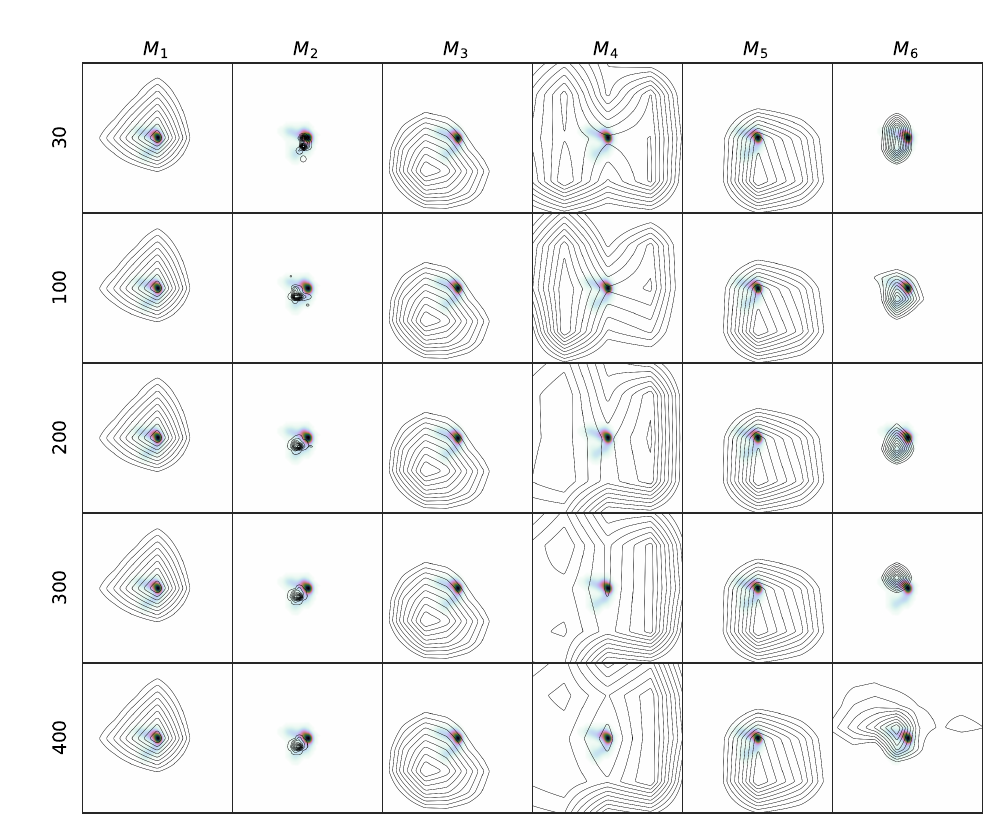}
    \caption{Evolution of Grad-CAM localisation contours with training epoch for a representative NAT source drawn from $\mathbf{R}_{L1}$. Rows correspond to the five checkpoints saved during training (30, 100, 200, 300, and 400 epochs, top to bottom); columns correspond to the six models $M_1$--$M_6$ (see Table~\ref{tab:models}). In each panel the background heatmap is the original radio source rendered with the cubehelix colourmap, and the overlaid iso-contours show the model's Grad-CAM activation; tightly nested, higher-level contours mark stronger activation. Reading down a column traces how each architecture's attention sharpens, drifts, or destabilises as training progresses, complementing the aggregate SSIM and entropy curves of Fig.~\ref{fig:ssim-H}. $M_2$ converges to contours that bend with the NAT tails and remain stable from epoch~30 onward; the CNN baselines $M_1$, $M_3$, and $M_5$ retain smooth, source-enclosing patterns with little epoch-to-epoch change; $M_4$ (Swin-B) keeps a broadly diffuse footprint at all checkpoints; and $M_6$ (ViT-B/16) visibly oscillates between a compact ring around the brightest pixel and a diffuse halo, mirroring the SSIM degradation.}
    \label{fig:loc-epochs}
\end{figure}

\end{appendix}
%%%% End of aa.dem

\end{document}